\documentclass[aip,jcp,onecolumn,superscriptaddress]{revtex4-1}%
\usepackage{epsfig,amssymb,amsmath,amsthm,amsfonts,amsbsy,mathrsfs}
\usepackage{graphicx,color}
\usepackage{algorithm}
\usepackage[noend]{algpseudocode}
\usepackage{epstopdf}
\usepackage{subcaption}
\usepackage{hyperref}

\newcommand{\bvec}[1]{\mathbf{#1}}

\newcommand{\Or}{\mathcal{O}}
\renewcommand{\Im}{\mathrm{Im}~}
\newcommand{\Tr}{\mathrm{Tr}}

\newcommand{\mc}[1]{\mathcal{#1}}

\newcommand{\ie}{\textit{i.e.}{}}

\newcommand{\abs}[1]{\left\lvert#1\right\rvert}
\newcommand{\norm}[1]{\left\lVert#1\right\rVert}

\renewcommand{\Im}{\mathfrak{Im}}

\newcommand{\vrr}{\bvec{r}}

\newcommand{\wt}[1]{\widetilde{#1}}

\begin{document}
\title{Robust Determination of the Chemical Potential in
the Pole Expansion and Selected Inversion Method for Solving Kohn-Sham
density functional theory}

\author{Weile Jia}
\thanks{jiaweile@berkeley.edu}
\affiliation{Department of Mathematics, University of California,
Berkeley, CA 94720, USA}

\author{Lin Lin}
\thanks{linlin@math.berkeley.edu}
\affiliation{Department of Mathematics, University of California,
Berkeley, CA 94720, USA} \affiliation{Computational Research
Division, Lawrence Berkeley National Laboratory, Berkeley, CA 94720,
USA}

\begin{abstract}
  Fermi operator expansion (FOE) methods are powerful alternatives to
  diagonalization type methods for solving Kohn-Sham density functional
  theory (KSDFT). One example is the pole expansion and selected
  inversion (PEXSI) method, which approximates the Fermi operator by
  rational matrix functions and reduces the computational complexity to
  at most quadratic scaling for solving KSDFT. Unlike diagonalization
  type methods, the chemical potential often cannot be directly read off
  from the result of a single step of evaluation of the Fermi operator.
  Hence multiple evaluations are needed to be sequentially performed to
  compute the chemical potential to ensure the correct number of
  electrons within a given tolerance. This hinders the performance of
  FOE methods in practice.  In this paper we develop an efficient and
  robust strategy to determine the chemical potential in the context of
  the PEXSI method. The main idea of the new method is not to find the
  exact chemical potential at each self-consistent-field (SCF) iteration
  iteration, but to dynamically and rigorously update the upper and
  lower bounds for the true chemical potential, so that the chemical
  potential reaches its convergence along the SCF iteration. Instead of
  evaluating the Fermi operator for multiple times sequentially, our
  method uses a two-level strategy that evaluates the Fermi operators in
  parallel.  In the regime of full parallelization, the wall clock time
  of each SCF iteration is always close to the time for one single
  evaluation of the Fermi operator, even when the initial guess is far away from the converged
  solution.  We demonstrate the effectiveness of the new method using
  examples with metallic and insulating characters, as well as results
  from \textit{ab initio} molecular dynamics.  
\end{abstract}

\maketitle

\section{Introduction}\label{sec:intro}

Kohn-Sham density functional theory (KSDFT) is the most widely used
theory for electronic-structure calculations. In the framework of
the self-consistent field (SCF) iteration, the
computational cost for solving KSDFT is mainly determined by the 
cost associated with the evaluation of the electron density for
a given Kohn-Sham potential during each iteration.  The
most widely used method to perform such an evaluation is to partially or
fully diagonalize the Kohn-Sham Hamiltonian, by means of computing a set of
eigenvectors corresponding to the algebraically smallest eigenvalues
of the Hamiltonian. The complexity of this approach is
$\mathcal{O}(N_e^3)$, where $N_e$ is the number of electrons in the
atomistic system of interest.  As the number of atoms or electrons in
the system increases, the cost of this diagonalization step becomes
prohibitively expensive.

In the past two decades, various numerical algorithms have been
developed for solving KSDFT without invoking the diagonalization
procedure. One particular
class of algorithms are the linear scaling 
algorithms~\cite{BowlerMiyazakiGillan2002,FattebertBernholc2000,HineHaynesMostofiEtAl2009,Yang1991,LiNunesVanderbilt1993,McWeeny1960,Goedecker1999,BowlerMiyazaki2012},
which relies on the near-sightedness principle for insulating systems
with large gaps~\cite{Kohn1996,ProdanKohn2005} to truncate elements of
the density matrix away from the diagonal. Among such methods, the Fermi operator expansion
method (FOE)~\cite{Goedecker1993} was also originally proposed as a linear
scaling method for insulating systems. Recently, the pole expansion and selected inversion
(PEXSI) method, which can be viewed as a FOE method by approximating the
Fermi operator using rational matrix functions, first achieved
computational complexity that is at most $\Or(N_{e}^2)$ for both
insulating and metallic
systems~\cite{LinLuYingE2009,LinChenYangEtAl2013,JacquelinLinYang2016}.
More specifically, the computational complexity of PEXSI depends on the
dimensionality of the system: the cost for quasi-1D systems such as
nanotubes is $\Or(N_{e})$ \ie{}  linear scaling; for quasi-2D systems
such as graphene and surfaces (slabs) is $\Or(N_{e}^{1.5})$; for general
3D bulk systems is $\Or(N_{e}^2)$.  PEXSI can be accurately applied to
general materials system including small gapped systems and metallic
systems, and remains accurate at low temperatures.  The PEXSI method has
a two-level parallelism structure and is by design highly scalable using
$10,000\sim 100,000$ processors on high performance machines.  The PEXSI
software package~\cite{PEXSI} has been integrated into a number of electronic
structure software
packages such as
BigDFT~\cite{MohrRatcliffBoulangerEtAl2014},
CP2K~\cite{VandeVondeleKrackMohamedEtAl2005}, 
SIESTA~\cite{LinGarciaHuhsEtAl2014,SolerArtachoGaleEtAl2002},
DGDFT~\cite{LinLuYingE2012,HuLinYang2015a},
FHI-aims~\cite{BlumGehrkeHankeEtAl2009}, QuantumWise ATK~\cite{ATK}, and has been used for
accelerating materials simulation with more than $10000$ atoms~\cite{HuLinYangEtAl2014,HuLinYangEtAl2016}.

One challenge for the PEXSI method, and for FOE methods using rational
approximations in general, is to determine the chemical potential $\mu$
so that the computed number of electrons at each SCF iteration is
$N_{e}$ within some given tolerance. This amounts to solving a scalar
equation
\begin{equation}
  N_{\beta}(\mu) = N_{e}.
  \label{eqn:Nerequire}
\end{equation}
Here $N_{\beta}(\mu)$ is the number of electrons evaluated using PEXSI at a given
chemical potential, and $\beta=1/(k_{B}T)$ is the inverse temperature.

Note that $N_{\beta}(\mu)$ is a monotonically non-decreasing function
with respect to $\mu$, and the simplest strategy to solve
Eq.~\eqref{eqn:Nerequire} is the bisection method.  However, starting
from a reasonably large search interval, the bisection method can require
tens of iterations to converge. This makes it more difficult to reach
the crossover point compared to diagonalization type methods, and
hinders the effectiveness of the method.  It also introduces potentially
large fluctuation in terms of the running time among different SCF
iterations.  One option to accelerate the convergence of the bisection
method is to use Newton's method, which takes the derivative
information into account and is expected to converge within a few
iterations. However, the effectiveness of Newton's method
relies on the assumption that the derivative of $N_{\beta}(\mu)$ with
respect to $\mu$ does not vanish. This assumption fails whenever $\mu$ is
inside a band gap.  A possible remedy is to use a regularized
derivative, but this may instead reduce the convergence rate and
increase the number of PEXSI evaluations per SCF iteration. 
Furthermore, when the initial guess is far away from the
true chemical potential, the derivative information is not very useful
in general. 
A robust
algorithm needs to handle all cases efficiently, and finds the solution
within at most a handful of evaluations starting from a possible wild
initial guess.  Hence the seemingly innocent scalar
equation~\eqref{eqn:Nerequire} turns out to be not so easy in practice.

In Ref.~\onlinecite{LinGarciaHuhsEtAl2014}, we have proposed a hybrid
Newton type method for determining the chemical potential in the context
of the PEXSI method. This method
uses an inertia counting strategy to rapidly reduce the size of the
search interval for the chemical potential, starting from a large search
interval. When the search interval becomes sufficiently small, a Newton
type method is then used. With the help of the inertia counting
strategy, the number of Newton steps is in general small (usually no
more than $5$), and each Newton step amounts to one step of 
evaluation of the Fermi operator. As discussed above, the Newton step may still occasionally
over-correct the chemical potential due to the small but not vanishing
derivative information, and the correction needs to be
discarded when it exceeds a certain threshold value. In such a case, the
inertia counting procedure needs to be invoked again with some updated
searching criterion. In this unfavorable regime, the effectiveness of
the PEXSI method is hindered. This procedure also inevitably
introduces extra tunable parameters, of which the values may be
difficult to predict \textit{a priori}.

In this work, we develop a new method for determining the chemical
potential that simultaneously improves the robustness and the efficiency
of the PEXSI method. Our main idea is to relax the requirement of
satisfying Eq.~\eqref{eqn:Nerequire} for each SCF iteration, so that the
error of the chemical potential is comparable to the residual error in
the SCF iteration, and the number of evaluations of the Fermi operator can therefore be
reduced. In particular, when the SCF converges, the chemical potential
converges as well and there is no loss of accuracy. The key to achieve this goal is to maintain 
rigorous lower and upper bounds for true chemical potential at each SCF
step, and to dynamically update the interval along the progress of the
SCF iteration. Due to the availability of these rigorous bounds, the new
method does not suffer from the over-correction problem as in Newton
type methods.  In the regime of full parallelization, the wall
clock time of the new method during each SCF iteration is always approximately
the same as that for one single evaluation of the Fermi operator, even when the initial
guess of the chemical potential is far away from the true solution.

The simplicity provided by the rigorous bounds of the chemical
potential also reduces the number of tunable parameters
in the practical implementation of the PEXSI method.  We find that the
remaining parameters are much less system dependent, and the default
values are already robust for both insulating and metallic systems. This
facilitates the usage of the PEXSI package as a black box software
package for general systems, and is therefore more user-friendly. 
We also compare the pole expansion obtained by an contour integral
approach~\cite{LinLuYingE2009}, and that obtained from an optimization
procedure by Moussa recently~\cite{Moussa2016}. We find that Moussa's
method can reduce the number of poles by a factor of $2\sim 3$ compared
to the contour integral approach, and hence we adopt this method as
the default option for pole expansion. 
Our new method is implemented in
the PEXSI software package~\cite{PEXSI}. For electronic structure
calculations, the PEXSI method can be accessed more easily from the
recently developed ``Electronic Structure Infrastructure'' (ELSI)
software package~\cite{ELSI}.  We demonstrate the performance of the
new method using the discontinuous Galerkin DFT (DGDFT) software
package~\cite{LinLuYingE2012,HuLinYang2015a},
using graphene and phospherene as
examples for metallic and insulating systems, respectively. Even though
the chemical potential is not fully accurate in each step of the SCF
iteration in the new method, we find that the number of SCF iterations
required to converge is almost the same as that needed by full
diagonalization methods, as well as the method in
Ref.~\onlinecite{LinGarciaHuhsEtAl2014} which requires multiple PEXSI
evaluations per SCF iteration. The accuracy of the new method is further
confirmed by \textit{ab initio} molecular dynamics, where we observe
negligible energy drift in an NVE simulation for a graphene system.

The rest of the paper is organized as follows. After reviewing the PEXSI
method in section~\ref{sec:pexsi}, we describe the new method for robust
determination of the chemical potential in section~\ref{sec:parallelmu}.
The numerical results are given in section~\ref{sec:results}, followed
by conclusion and discussion in section~\ref{sec:conclusion}. Some
estimates related to the update of the bounds among consecutive
SCF steps is given in Appendix~\ref{app:eigenvalue}.

\section{Pole expansion and selected inversion method}\label{sec:pexsi}

Assume the Kohn-Sham orbitals are expanded with a set of basis functions
$\Phi=[\varphi_1,\cdots,\varphi_{N}]$, and denote by $H,S$ the discretized Kohn-Sham
Hamiltonian matrix and overlap matrices, respectively. Without loss of
generality we consider isolated systems or solids with $\Gamma$ point
sampling strategy of the Brillouin zone, and $H,S$ are real symmetric
matrices. The standard method for solving the discretized system is to
solve the generalized eigenvalue problem 
\begin{equation}
  H C = S C \Lambda
  \label{eqn:geneig}
\end{equation}
via direct or iterative methods. Here $\Lambda =
\mathrm{diag}[\lambda_{1},\ldots,\lambda_{N}]$ is a diagonal matrix
containing the Kohn-Sham eigenvalues.
The single particle density matrix is then defined as
\begin{equation}
  \Gamma = C f_{\beta}(\Lambda-\mu I) C^{T}.
  \label{eqn:DM}
\end{equation}
Here $\beta=1/k_{B} T$ is the inverse temperature, and
\begin{equation}
f_{\beta} (x) = \frac{1}{1+e^{\beta x}}
\label{eqn:fermidirac}
\end{equation}
is the Fermi-Dirac distribution (spin degeneracy is omitted).
The chemical potential $\mu$ chosen to ensure that
\begin{equation}
  N_{\beta}(\mu)=\Tr[f_{\beta}(\Lambda-\mu I)] =  \Tr[S \Gamma] = N_{e},
\label{eqn:normalize}
\end{equation}
where $N_{e}$ is the number of electrons.
The computational complexity for the solution of the generalized
eigenvalue problem is typically $\Or(N_{e}^{3})$.

Since $f_{\beta}(\cdot)$ is a smooth function when $\beta$ is finite,
the Fermi operator expansion (FOE) method expands $f_{\beta}(\cdot)$ 
using a linear combination of simple functions such as polynomials or rational functions, so that the
density matrix can be evaluated by matrix-matrix multiplication or
matrix inversion, without computing any eigenvalue or eigenfunction.
The recently
developed pole expansion and selected inversion (PEXSI)
method~\cite{LinLuYingE2009,LinChenYangEtAl2013} is one type of FOE
methods, which expands $\Gamma$ using a rational approximation as
\begin{equation}
  \Gamma \approx  \sum_{l=1}^{P}\Im \left(\omega^{\rho}_l (H - (z_l+\mu)
  S)^{-1}\right).
  \label{eqn:poleexpandDM}
\end{equation}
Here $G_{l}:=(H - (z_l+\mu) S)^{-1}$ defines a inverse matrix (or
Green's function) corresponding to the complex shift $z_{l}$. 
The pole expansion is not unique. The expansion in Ref.~\onlinecite{LinLuYingE2009} 
uses a
discretized Cauchy contour integral technique, and can accurately
approximate the density matrix with only $\Or(\log \beta\Delta E)$
terms. Here $\Delta E:=\max_{1\le i\le N}|\mu-\lambda_{i}|$ is the spectral
radius of the matrix pencil $(H,S)$.  The number of poles needed in
practice is typically $40\sim 80$. 

Recently, a new expansion for the Fermi-Dirac
function has been developed by Moussa~\cite{Moussa2016}, which has the same form as in
Eq.~\eqref{eqn:poleexpandDM} but with a different choice of complex
shifts $\{z_{l}\}$ and weights $\{\omega_{l}\}$. This approach is based
on modifying the Zolotarev expansion for the sign function~\cite{}
through numerical optimization. We find that this new approach further reduces the number of
poles to $10\sim 30$ for approximating the Fermi-Dirac function for a
wide range of $\beta\Delta E$. 
Furthermore, the number of poles only approximately depends on $\beta
\Delta E_{o}$. Here $\Delta E_{o}:=\max_{1\le i\le N}(\mu -
\lambda_{i})$ is the spectral radius corresponding to the occupied
eigenvalues, which can be significantly smaller than the spectral radius
of the matrix pencil $(H,S)$. Due to these advantages, we adopt Moussa's
method as the default option for the pole expansion for the density
matrix.  

However, the optimization based pole expansion approach has two
minor drawbacks compared to the contour integral approach.  First, the
validity of the contour integral approach only depends on the analytical
structure of the function to be approximated in the complex plane,
rather than the detailed form of the function. This fact allows us to
use exactly the same set of poles to approximate multiple matrix
functions simultaneously, such as the energy density matrix and the free
energy density matrix used for computing the Pulay force and the
electronic entropy,
respectively~\cite{LinChenYangEtAl2013,LinGarciaHuhsEtAl2014}.  This
property does not hold for optimization based pole expansion method. 
Second, the contour integral approach is a semi-analytic approach, and
can achieve very high accuracy (e.g. the error of the force can be as
small as $10^{-9}$ Hartree/Bohr when compared to results from
diagonalization method) without suffering from numerical problems. On
the other hand, the optimization problem for finding the pole expansion
can become increasingly ill-conditioned as the requirement of accuracy
increases. 

In order to obtain the electron density in the real space, it is not
necessary to evaluate the entire density matrix $\Gamma$. When the
local or semilocal exchange-correlation functionals are used, 
only the {\em selected elements} $\{\Gamma_{ij}\vert H_{ij}\ne 0\}$ are
in general needed, even if the off-diagonal elements of $\Gamma$ decay
slowly as in the case of metallic systems. We remark that when a matrix
element $H_{ij}$ is accidentally zero (often due to symmetry
conditions), such zero elements should also be treated as nonzero
elements. According to Eq.~\eqref{eqn:poleexpandDM}, we only need to
evaluate these selected elements of each Green's function.  This can be
achieved via the selected inversion
method~\cite{LinLuYingE2009,LinYangMezaEtAl2011,JacquelinLinYang2016}.
For a (complex) symmetric matrix of the form $A=H-zS$, the selected
inversion algorithm first constructs an $LDL^T$ or $LU$ factorization of $A$.
The computational scaling of the
selected inversion algorithm is only related to the number of
nonzero elements in the $L,U$ factors. The computational
complexity is $\Or(N)$ for
quasi-1D systems, $\Or(N^{1.5})$ for quasi-2D systems, and $\Or(N^{2})$
for 3D bulk systems, thus achieving universal asymptotic improvement
over the diagonalization method for systems of all dimensions.  It
should be noted that the selected inversion algorithm is an
\textit{exact} method for computing selected elements of $A^{-1}$ if
exact arithmetic is employed, and in practice the only source of error
originates from the roundoff error.  In particular, the selected
inversion algorithm does not rely on any localization property of
$A^{-1}$.

In addition to its favorable asymptotic complexity, the
PEXSI method is also inherently more scalable than the standard
approach based on matrix diagonalization when it is implemented on 
a parallel computer.  The parallelism in PEXSI exists at two levels.
First, the $LU$ factorization and the selected inversion processes
associated with different poles are completely independent.  Second,
each $LU$ factorization and selected inversion can
be parallelized. We use the SuperLU\_DIST~\cite{LiDemmel2003} package
for parallel $LU$ factorization, and 
PSelInv~\cite{JacquelinLinYang2016,JacquelinLinWichmannEtAl2016}
for parallel selected inversion, respectively.
In practice the PEXSI method can harness over $100,000$ processors on
high performance computers.

\section{Robust determination of the chemical potential}\label{sec:parallelmu}

Unlike diagonalization type methods, in general the chemical potential
$\mu$ cannot be read off directly from one evaluation of the Fermi
operator. Typically
several iterations are needed to identify the chemical potential in
order to satisfy the equation~\eqref{eqn:normalize}. As discussed in the
introduction, the behavior of the function $N_{\beta}$ can depend on
whether the system is insulating or metallic, and whether the
temperature is low or high compared to the magnitude of the gap.
Furthermore, when the initial guess is far away from the true chemical
potential, the derivative information $N_{\beta}'(\mu)$ is in general
not very useful.  Figure~\ref{fig:Nbeta} illustrates that the behavior of
$N_{\beta}(\mu)$ for an insulating and a metallic system, respectively.
The details of the setup of the systems are  given in
section~\ref{sec:results}.

\begin{figure}[H]\label{fig:Nbeta}
\begin{subfigure}{.5\textwidth}
\includegraphics[width=0.7\textwidth,angle=270]{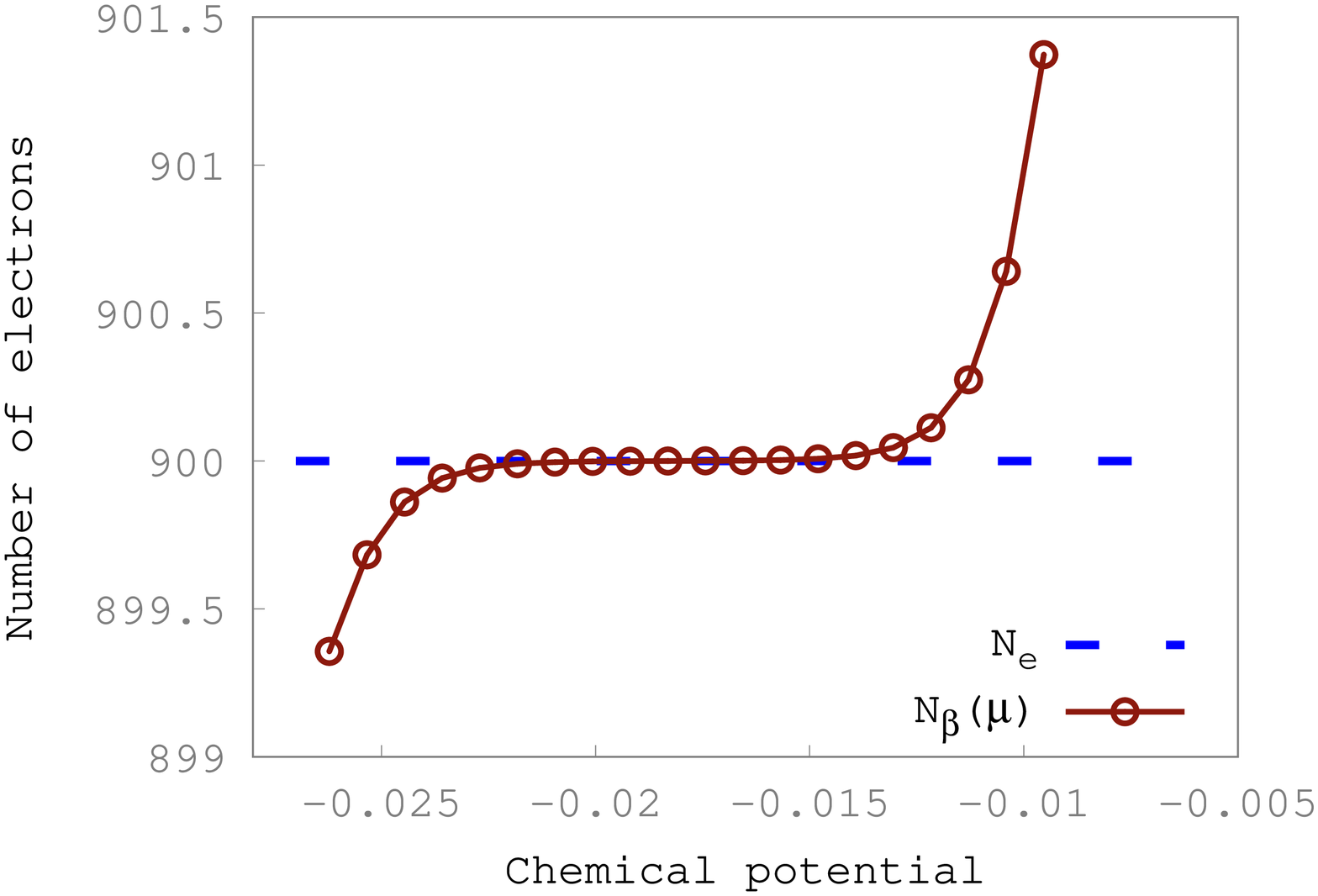}
\label{fig1:sub1}
\end{subfigure}
\begin{subfigure}{.5\textwidth}
\includegraphics[width=0.7\textwidth,angle=270]{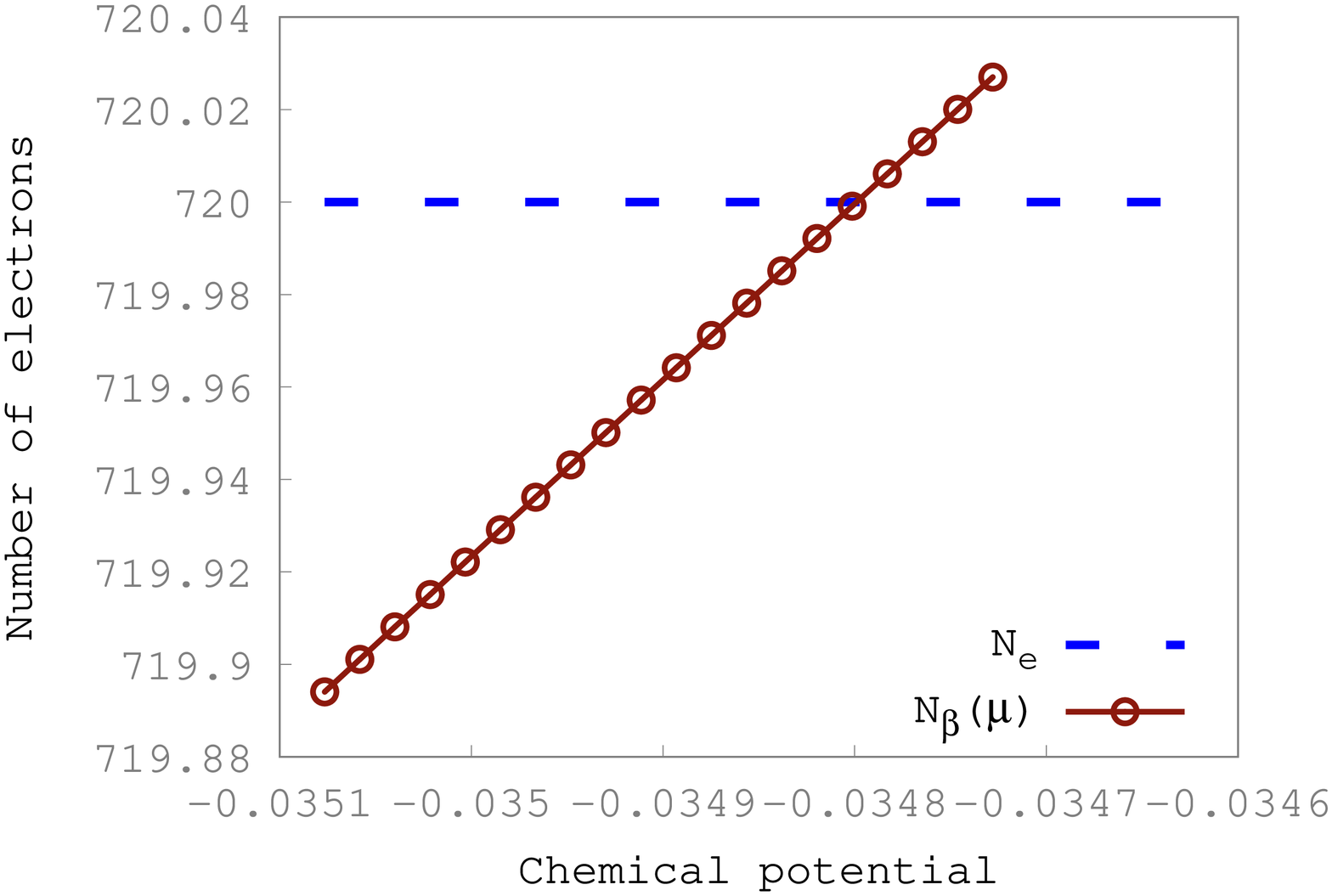}
\label{fig1:sub2}
\end{subfigure}
\caption{Fermi Dirac distribution for phospherene nanoribbon(PNR) 180 atoms(left) and graphene(GRN) 180 atoms (right).The blue dash line 
shows the exact number of electrons (900 for PNR and 720 for
GRN).
}
\label{fig:Nbeta}
\end{figure}


Instead of converging the chemical potential in each SCF iteration, 
the main idea of this paper is to compute rigorous upper and lower
bounds for the chemical potential. The chemical potential is obtained
using such bounds through an interpolation procedure, and may not
necessarily be fully accurate in each SCF iteration. However, as the SCF
iteration converges, the size of the search interval characterized by
the bounds decreases to zero, and hence the chemical potential also
converges to the true solution.

We obtain such bounds through a coarse level and a fine level procedure
as follows. At the coarse level, we use an inertia counting procedure
previously developed in Ref.~\onlinecite{LinGarciaHuhsEtAl2014}, which
is an inexpensive procedure to compute the zero temperature limit
$N_{\beta=\infty}(\mu)$ for a set of values of $\mu$.
Since the temperature effect is usually on the order of
$100$ K, which is orders of magnitude smaller than the size of the
search interval which is on the order of Hartree (1 Hartree $\approx
315774$ K), such zero temperature information can provide estimates of
the upper and lower bounds until the finite temperature effects becomes
non-negligible. Then at the fine level, we use PEXSI to evaluate
$N_{\beta}(\mu)$ for a smaller set of values of $\mu$ (the size of this
set is denoted by $N_{\text{point}}$), 
which properly takes the finite temperature effect into
account and refines the bounds. The evaluation of the Fermi operator at multiple values
of $\mu$ also allows us to interpolate the chemical potential as well as
the density matrix, so that Eq.~\eqref{eqn:Nerequire} is satisfied up to
the error of the interpolation procedure. The procedure above is all
that is involved in a single SCF iteration, and no further iteration is
necessary within the SCF iteration. 

At the coarse level and the fine level, the multiple evaluations of
$N_{\infty}(\mu)$ and $N_{\beta}(\mu)$ can be evaluated in
an embarrassingly parallel fashion. Compared to
Ref.~\onlinecite{LinGarciaHuhsEtAl2014}, this adds a third layer of
parallelism.
This is the key to achieve high efficiency using the PEXSI method. We
assume the total number of processors is denoted by
\begin{equation}
  N_{\text{proc}}=N_{\text{sparse}} N_{\text{pole}}  N_{\text{point}},
  \label{eqn:Nproc}
\end{equation}
where $N_{\text{sparse}}$ is the number of processors for operations on
each matrix, such as $LU$ factorization or selected inversion.
$N_{\text{pole}}=P$ is the number of poles in the pole expansion.  When
the number of processors is a multiple of $N_{\text{pole}}
N_{\text{point}}$, all poles can be evaluated fully in parallel, and we
refer to this case the full parallelization regime.  In this case the
wall clock time of the new method during each SCF iteration is
approximately the same as that for one single evaluation of the Fermi
operator.

In order to guarantee that $\mu$ can converge to the true value of the
chemical potential when SCF converges, the upper and lower bounds of the
chemical potential must also reduce proportionally with respect to the
residual error in the SCF iteration. Our method dynamically updates the
interval between consecutive SCF steps, which is rigorously controlled
by the magnitude of the change of the Kohn-Sham potential. In
particular, when the SCF iteration is close to its convergence, the size
of the search interval characterized by the upper and lower bounds is
smaller than the finite temperature effect. In such case, the coarse
level inertia counting procedure can be safely skipped, and the wall
clock time is precisely the same as that for one single PEXSI
evaluation.


\subsection{Coarse level: Inertia counting}

While the computation of $N_\beta(\mu)$ requires the 
evaluation of the Fermi operator, it turns out to be much easier to compute
$N_{\infty}(\mu)$ without diagonalizing the matrix pencil
$(H,S)$. Here the subscript $\beta=\infty$ refers to the zero
temperature limit. 
The method of Ref.~\onlinecite{LinGarciaHuhsEtAl2014} uses
the Sylvester's law of inertia~\cite{Sylvester1852}, which
states that the inertia (the number of negative, zero and positive
eigenvalues, respectively) of a real symmetric matrix does not change under a
congruent transform. 
Our strategy is to perform a matrix decomposition of the shifted matrix 
\[
H-\mu S = L D L^{T},
\]
where $L$ is unit lower triangular and $D$ is diagonal.  
Since $D$ is congruent to $H-\mu S$, $D$ has the same inertia as that of
$H-\mu S$. Hence, we can obtain $N_{\infty}(\mu)$ by simply
counting the number of negative entries in $D$.  
The matrix
decomposition can be computed efficiently
by using a sparse $LDL^{T}$ or $LU$ factorization with a symmetric
permutation strategy. 
The same
conclusion holds when $H$ is Hermitian, and in this case one then
replaces the $LDL^{T}$ factorization by the $LDL^{*}$ factorization,
where $L^{*}$ is the Hermitian conjugate of $L$.
Compared to the evaluation of the Fermi operator using PEXSI, the
inertia counting step is fast for a number of reasons: 1) PEXSI requires
both the sparse factorization and selected inversion, and inertia
counting only requires a sparse factorization. 2) PEXSI requires
evaluations of $P$ Green's functions to obtain one value of
$N_{\beta}(\mu)$, and inertia counting obtains $N_{\infty}(\mu)$ with
one factorization. 3) PEXSI requires complex arithmetic, and for real
matrices the inertia counting procedure only requires real arithmetic
and thus fewer floating point operations. Hence the inertia counting
step takes only a fraction of the time by each PEXSI evaluation.

The inertia counting strategy is naturally suited for
parallelization.  The $N_{\text{proc}}$ processors in
Eq.~\eqref{eqn:Nproc} can be
partitioned into $N_{\text{pole}}  N_{\text{point}}$ groups. 
Starting from an initial guess interval $(\mu_{\min},\mu_{\max})$,
a set of values $\{N_{\infty}(\mu_{g})\}$ 
can be simultaneously evaluated on a uniform grid
\begin{equation}
  \mu_{g} = \mu_{\min} + \frac{g-1}{N_{\text{pole}}N_{\text{point}}-1}(\mu_{\max}-\mu_{\min}),\quad
  g=1,\ldots,N_{\text{pole}}N_{\text{point}}.
  \label{eqn:muginertia}
\end{equation}
The inertia counting procedure can determine that only one of the interval
should contain the chemical potential, and the same procedure can be
applied to this refined interval. It is easy to see that the size of the
search interval shrinks rapidly as $(N_{\text{pole}}N_{\text{point}}-1)^{-k}$ with respect to the number of
iterations $k$. For example when $N_{\text{pole}}N_{\text{point}}=40$,
$3$ iterations reduces the search interval size from $10$ Hartree to $4$
meV.

The effectiveness of the inertia searching procedure relies on that
$N_{\beta}(\mu)$ can be well approximated by
$N_{\infty}(\mu)$. This approximation is clearly valid when the search
interval is much larger than $1/\beta=k_{B} T$. When the search interval
is comparable to $k_{B} T$, the difference of the two quantities becomes
noticeable, and care must be taken so that the search interval does not
become too small to leave the true chemical potential outside.  In
Ref.~\onlinecite{LinGarciaHuhsEtAl2014}, we employ an interpolation
procedure to estimate $N_{\beta}(\mu)$ from $N_{\infty}(\mu)$. The
potential drawback of this procedure is that the true chemical potential may not
be always included in the search interval. In this case, the subsequent
PEXSI step can fail, and one must go back to
the previous inertia counting stage with an expanded search interval.

\begin{figure}[H]
\begin{subfigure}{.5\textwidth}
\includegraphics[width=0.7\textwidth,angle=270]{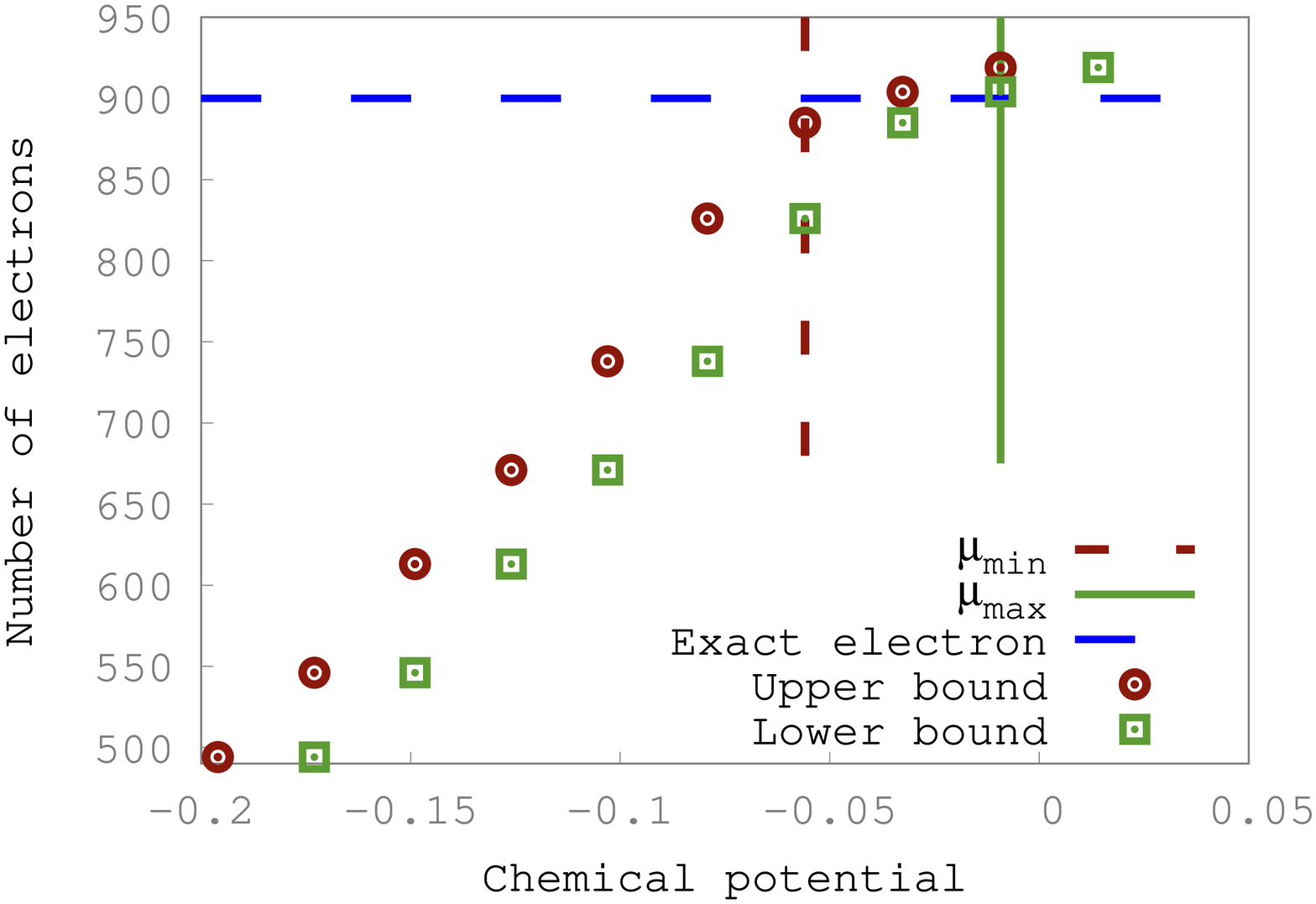}
\caption{Inertia counting step 1.}
\label{fig2:sub1}
\end{subfigure}
\begin{subfigure}{.5\textwidth}
\includegraphics[width=0.7\textwidth, angle=270]{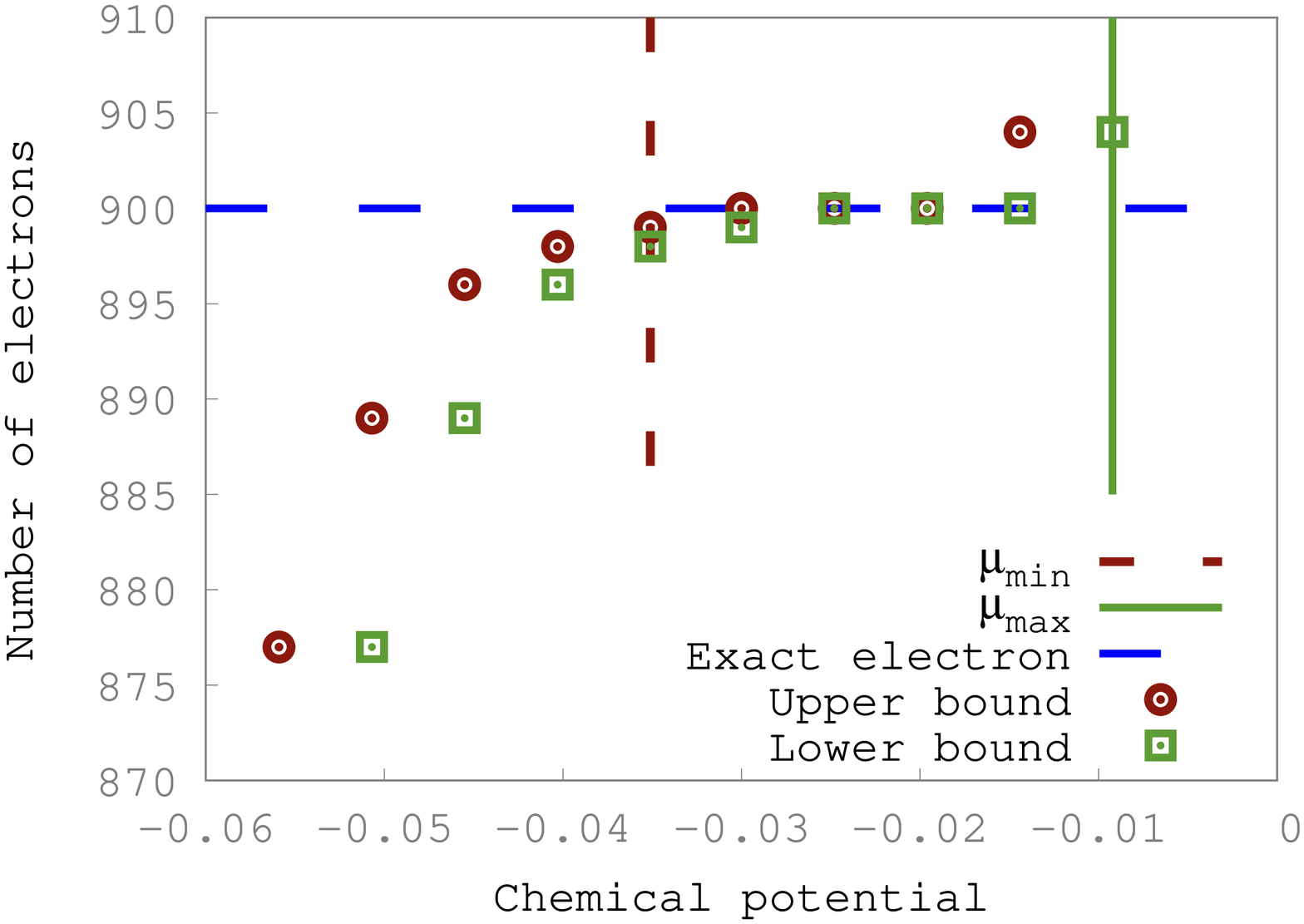}
\caption{Inertia counting step 2.}
\label{fig2:sub2}
\end{subfigure}
\begin{subfigure}{.5\textwidth}
\includegraphics[width=0.7\textwidth,angle=270]{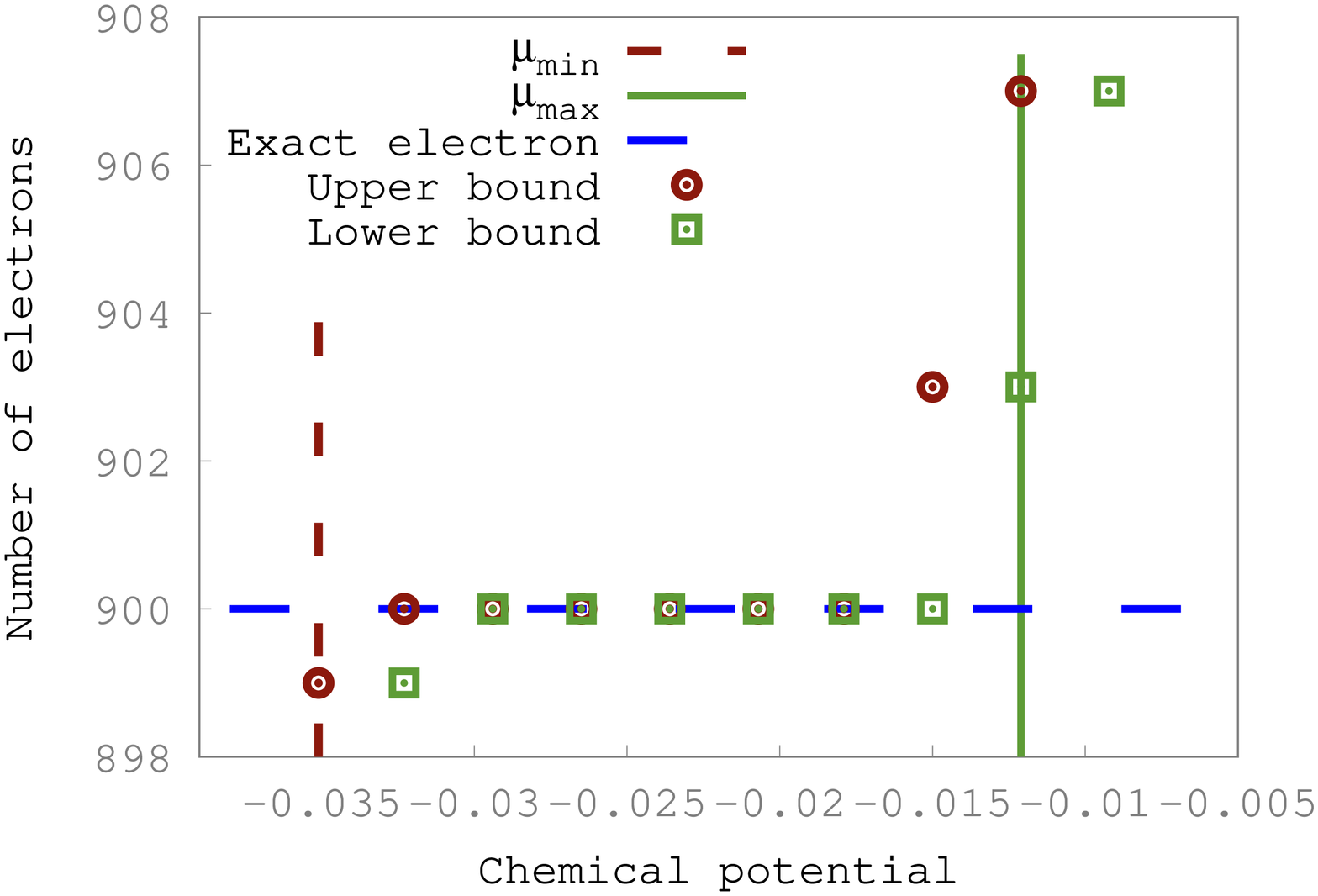}
\caption{Inertia counting step 3.}
\label{fig2:sub3}
\end{subfigure}
\begin{subfigure}{.5\textwidth}
\includegraphics[width=0.7\textwidth,angle=270]{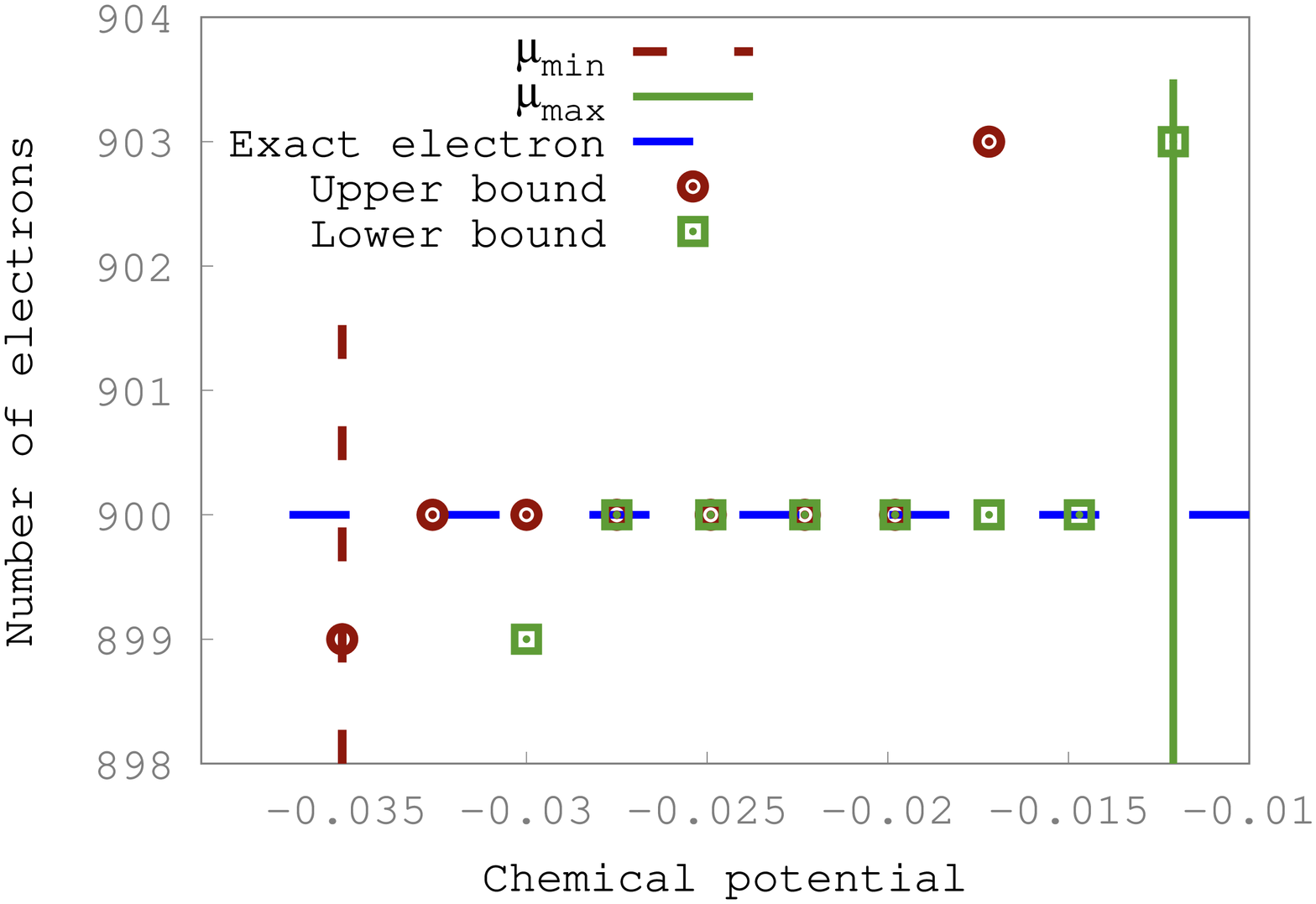}
\caption{Inertia counting step 4.}
\label{fig2:sub4}
\end{subfigure}
\caption{Refinement of the bounds for the chemical potential in 4 inertia 
counting steps for the PNR 180 system. 
For illustration purpose we evaluate the inertia counting on 10 points.
In step 4, the inertia counting  procedure
stops because the upper and lower bounds can not be further refined. 
}
\label{fig2}
\end{figure}

In this work, we use the information $N_{\infty}(\mu_{g})$ only to
calculate the upper and lower bounds for $N_{\beta}(\mu_{g})$, which guarantees that
the true chemical potential is always contained in the search interval,
up to the error controlled by a single parameter $\tau_{\beta}$.
More specifically, since the Fermi-Dirac function $f_{\beta}(x)$ is a
non-increasing function and rapidly approaches $1$ when $x<0$
and $0$ when $x>0$, we can select a number $\tau_{\beta}$ so
that we can approximate
\begin{equation}
  f_{\beta}(x) \approx \begin{cases}
    1, & x \le -\tau_{\beta},\\
    0, & x \ge  \tau_{\beta}.
  \end{cases}
  \label{eqn:fermiapprox}
\end{equation}
$\tau_{\beta}$ is a tunable parameter but is not system dependent. 
In practice we find
that setting $\tau_{\beta}=3/\beta = 3k_{B} T$ is a sufficiently
conservative value for the robustness of our method.  With this controlled approximation, we have
\begin{equation}
  N_{\beta}(\mu-\tau_{\beta}) \le N_{\infty}(\mu)
  \le N_{\beta}(\mu+\tau_{\beta}).
  \label{eqn:NbetaEst}
\end{equation}
Hence each evaluation of $N_{\infty}$ provides an upper and a lower
bound for $N_{\beta}$ at two other energy points according
to~\eqref{eqn:NbetaEst}. 
This provides an estimate of the chemical potential on each grid point
of the uniform grid in the inertia counting. If $\mu-\tau_{\beta}$
exceeds the lower bound or $\mu+\tau_{\beta}$ exceeds the upper bound,
we can just take the lower bound to be $0$, or the upper bound to be the
matrix size, respectively. We set $\mu_{\min}$ to be the largest $\mu$
so that the upper bound is below $N_{e}$, and $\mu_{\max}$ is the
smallest $\mu$ so that the lower bound is above $N_{e}$.
Fig.~\ref{fig2} illustrates the refinement procedure of the bounds for the
chemical potential for the PNR 180 system.


The inertia counting stops when $\mu_{\max}-\mu_{\min}$ is below certain
tolerance denoted by $\tau^{\mu}_{\mathrm{inertia}}$, or when the lower and upper bound
cannot be further refined, whichever is satisfied first. In particular, if the size of the initial
search interval is smaller than the given tolerance, the inertia
counting procedure should be skipped directly. As will be discussed
later, the capability of skipping the inertia counting procedure is
important for the self-correction of the search interval for the
chemical potential. The output of the inertia counting procedure is an
updated search interval, still denoted by $(\mu_{\min},\mu_{\max})$
ready for the evaluation of the Fermi operator.

\subsection{Fine level: Multiple point evaluation of the Fermi operator}\label{subsec:pexsimu}


After the inertia counting step converges, we refine the upper and lower
bounds of the chemical potential by evaluating the Fermi operator on
multiple points using PEXSI. This step also gives the density matrix $\Gamma$.
Our target is to minimize the
\textit{wall clock time} with the help of parallelization. Hence
motivated from the inertia counting procedure that performs multiple
matrix factorizations simultaneously, we can also perform
$N_{\text{point}}$ PEXSI evaluations simultaneously. 
In the current context, the method in
Ref.~\onlinecite{LinGarciaHuhsEtAl2014} can be regarded as choosing
$N_{\text{point}}=1$.

Starting from the search interval $(\mu_{\min},\mu_{\max})$ from the
output of the inertia counting procedure, this interval is uniformly
divided into $N_{\text{point}}-1$ intervals as
\begin{equation}
  \mu_{g} = \mu_{\min} + \frac{g-1}{N_{\text{point}}-1}(\mu_{\max}-\mu_{\min}),\quad
  g=1,\ldots,N_{\text{point}},
  \label{eqn:mugpexsi}
\end{equation}
and the density matrix $\Gamma(\mu_{g})$ and hence number of electrons
$N_{\beta}(\mu_{g})$ can be evaluated simultaneously for all points.
Unlike inertia counting, $N_{\beta}(\mu_{g})$ is evaluated accurately
with the finite temperature effect correctly taken into account, and the
only error is from the pole expansion.  This naturally updates the upper
and lower bound of the chemical potential $\mu_{\max},\mu_{\min}$,
respectively.  If $\abs{N_{\beta}(\mu_{g})-N_{e}}$ is already smaller
than the given tolerance $\tau^{N_{e}}$ for some $g$, then the PEXSI
step converges. We set $\mu=\mu_{g},\Gamma=\Gamma(\mu_{g})$ and we may
proceed to the next SCF iteration.  If the condition is satisfied for
multiple $g$, we simply choose the first $\mu_{g}$ that satisfies the
convergence criterion.
If the convergence criterion is not met for any $\mu_{g}$, we construct a piecewise interpolation polynomial
$\wt{N}_{\beta}(\mu)$ that is monotonically non-decreasing, and
satisfies
\begin{equation}
  \wt{N}_{\beta}(\mu_{g}) = N_{\beta}(\mu_{g}),\quad g=1,\ldots,N_{\text{point}}.
  \label{}
\end{equation}
Then the chemical potential is identified by solving
$\wt{N}_{\beta}(\mu)=N_{e}$. Such an interpolation can be constructed
using e.g. the monotone cubic spline interpolation~\cite{Wolberg1999}. In
practice we find the following linear interpolation procedure is simpler
and works almost equally efficiently: We identify an interval
$(\mu_{g^{*}},\mu_{g^{*}+1})$ so that
$N_{\beta}(\mu_{g^{*}})<N_{e}<N_{\beta}(\mu_{g^{*}+1})$.  
Then the chemical
potential is found by solving the linear equation
\begin{equation}
  N_{e} = N_{\beta}(\mu_{g}^{*}) + 
  \frac{\mu-\mu_{g^{*}}}{\mu_{{g}^{*}+1}-\mu_{g^{*}}}
  \left( N_{\beta}(\mu_{{g}^{*}+1}) - N_{\beta}(\mu_{g}^{*})\right).
  \label{eqn:linearmixmu}
\end{equation}
Once $\mu$ is obtained, the density matrix is linearly mixed similarly as
\begin{equation}
  \Gamma = \Gamma(\mu_{g}^{*}) + 
  \frac{\mu-\mu_{g^{*}}}{\mu_{{g}^{*}+1}-\mu_{g^{*}}}
  \left( \Gamma(\mu_{{g}^{*}+1}) - \Gamma(\mu_{g}^{*})\right).
  \label{eqn:linearmixgamma}
\end{equation}

Note that the interpolation procedure does not guarantee that $\mu$
satisfies the condition~\eqref{eqn:normalize}. However, it will ensure
that the search interval is reduced at least by a factor
$(N_{\text{point}}-1)^{-1}$, and hence with $k$ steps of iteration, the 
convergence rate of the chemical potential is at least $(N_{\text{point}}-1)^{-k}$.
At each step of the iteration, the true chemical potential
is always contained in the search interval. Hence the refinement is more
robust than Newton type methods, and it does not need to fold back to
the inertia counting stage.  The extra flexibility in choosing
$N_{\text{point}}$ means that faster convergence can be achieved when a
larger amount of computational resource is available. We also remark
that the two end points of the search interval in
Eq.~\eqref{eqn:mugpexsi} are always the lower and upper bounds for
$\mu$, and hence can often be discarded in practical calculations. This
increases the efficiency especially when $N_{\text{point}}$ is small. In
practice we find that even the extreme case with $N_{\text{point}}=2$
with dropped end points is already very robust, and this is the default
choice in our implementation.

\subsection{Dynamical update of the search interval}\label{subsec:correctmu}

%

After the inertia counting and the multiple point evaluation step,
the search interval $(\mu_{\min},\mu_{\max})$ already provides
very accurate information of the upper and lower bound for the chemical
potential. Here we demonstrate how to reuse this information in the
following SCF iteration, while maintaining rigorous bounds for the
chemical potential of the updated Hamiltonian matrix.

For KSDFT calculations with local and semi-local
exchange-correlation functionals, the change of the Kohn-Sham
Hamiltonian during the SCF iterations is always given by a local
potential denoted by $\Delta V_{\text{SCF}}(\vrr)$.  Define
\begin{equation}
  \Delta V_{\min}=\inf_{\vrr} \Delta V_{\text{SCF}}(\vrr), \quad \Delta
  V_{\max}=\sup_{\vrr} \Delta V_{\text{eff}}(\vrr). 
  \label{}
\end{equation}
Then following the derivation in
Appendix~\ref{app:eigenvalue}, the chemical potential must be contained
in the interval $(\mu_{\min}+\Delta V_{\min},\mu_{\max}+\Delta
V_{\max})$. This interval provides an upper and
lower bound for the chemical potential for the new matrix pencil
$(H,S)$, and can be used as the initial search interval
in the next SCF iteration.

Note that when $(\mu_{\max}-\mu_{min})+(\Delta V_{\max}-\Delta
V_{\min})$ is smaller than the stopping criterion of the inertia
counting $\tau_{\beta}$, the inertia counting step will be skipped
automatically and only the PEXSI step will be executed. When the SCF
iteration is close to its convergence,  $\Delta V_{\max}-\Delta V_{\min}$ becomes
small, and the search interval will be systematically reduced.  
A pseudocode for our method is summarized in Alg.~\ref{alg:robust}.

\begin{algorithm}[H]
%
%
%
%
\leftline{\textbf{Input:}}
$\tau^{\mu}_{\mathrm{inertia}},\tau^{N_{e}}$, $N_{\text{proc}}=N_{\text{sparse}}
N_{\text{point}} N_{\text{pole}}$ (assuming full parallelization), 

$(\mu_{\min},\mu_{\max})$ as the initial search interval.


%
%
%

\leftline{\textbf{Output:}}

Converged density matrix $\Gamma$ and chemical potential $\mu$.

\begin{algorithmic}[1]
  \While {SCF has not converged}
  \State {Construct the projected Hamiltonian matrix $H$ and overlap
  matrix $S$}

  \While {Inertia counting has not converged} 
  \For {$g=1$ to $M N_{\text{point}}$} 
  \State{Evaluate $N_{\infty}(\mu_{g})$ for
  the processor group associated with $\mu_{g}$ 
  from Eq.~\eqref{eqn:muginertia}.
  }
\EndFor
\State{Construct the upper and lower bounds for
$N_{\beta}(\mu_{g})$ for each point $\mu_{g}$.}
\State Update the search interval $(\mu_{\min},\mu_{\max})$.
    \EndWhile

    \For {$g=1$ to $N_{\text{point}}$}
    \For {$l=1$ to $N_{\text{pole}}$}
    \State{Evaluate the pole $(H - (z_l+\mu)S)^{-1}$ for the 
    processor group associated with $l$
    }
    \EndFor
    \State{
    Evaluate $\mu$ and $\Gamma$ for
    processor group associated with $\mu_{g}$ from 
    Eq.~\eqref{eqn:mugpexsi} if needed.
    }
  \EndFor

  \State{Perform linear interpolation for 
  $\mu$ and 
  $\Gamma$ according to
  Eq.~\eqref{eqn:linearmixmu} and~\eqref{eqn:linearmixgamma}.}
  \State{Evaluate the density and the new potential. Compute the
  difference of the potential $\Delta V$.}
  \State {Update the search interval $(\mu_{\min},\mu_{\max})\gets 
  \left( \mu_{\min} + \Delta V_{\min}, \mu_{\max} + \Delta
  V_{\max}\right)$}.
\EndWhile
\end{algorithmic}
\caption{Robust determination of the chemical potential.}
\label{alg:robust}
\end{algorithm}

\section{Numerical Results}
\label{sec:results}

We have implemented the new method in the PEXSI package~\cite{PEXSI},
which is a standalone software package for evaluating the density matrix
for a given matrix pencil $(H,S)$. For electronic structure
calculations, the PEXSI package is also integrated into the recently
developed ``Electronic Structure Infrastructure'' (ELSI)
software package~\cite{ELSI}.  In order to demonstrate that the
performance of the new
method in the context of SCF iterations for real materials, we test its performance
in the DGDFT (Discontinuous Galerkin Density Functional
Theory) software package~\cite{LinLuYingE2012,HuLinYang2015a}, which is
a massively parallel electronic structure software package for large
scale DFT calculations using adaptive local basis functions (ALB). 

\begin{figure}[H]
\centering
\begin{subfigure}{.45\textwidth}
\includegraphics[width=0.7\textwidth, angle=270]{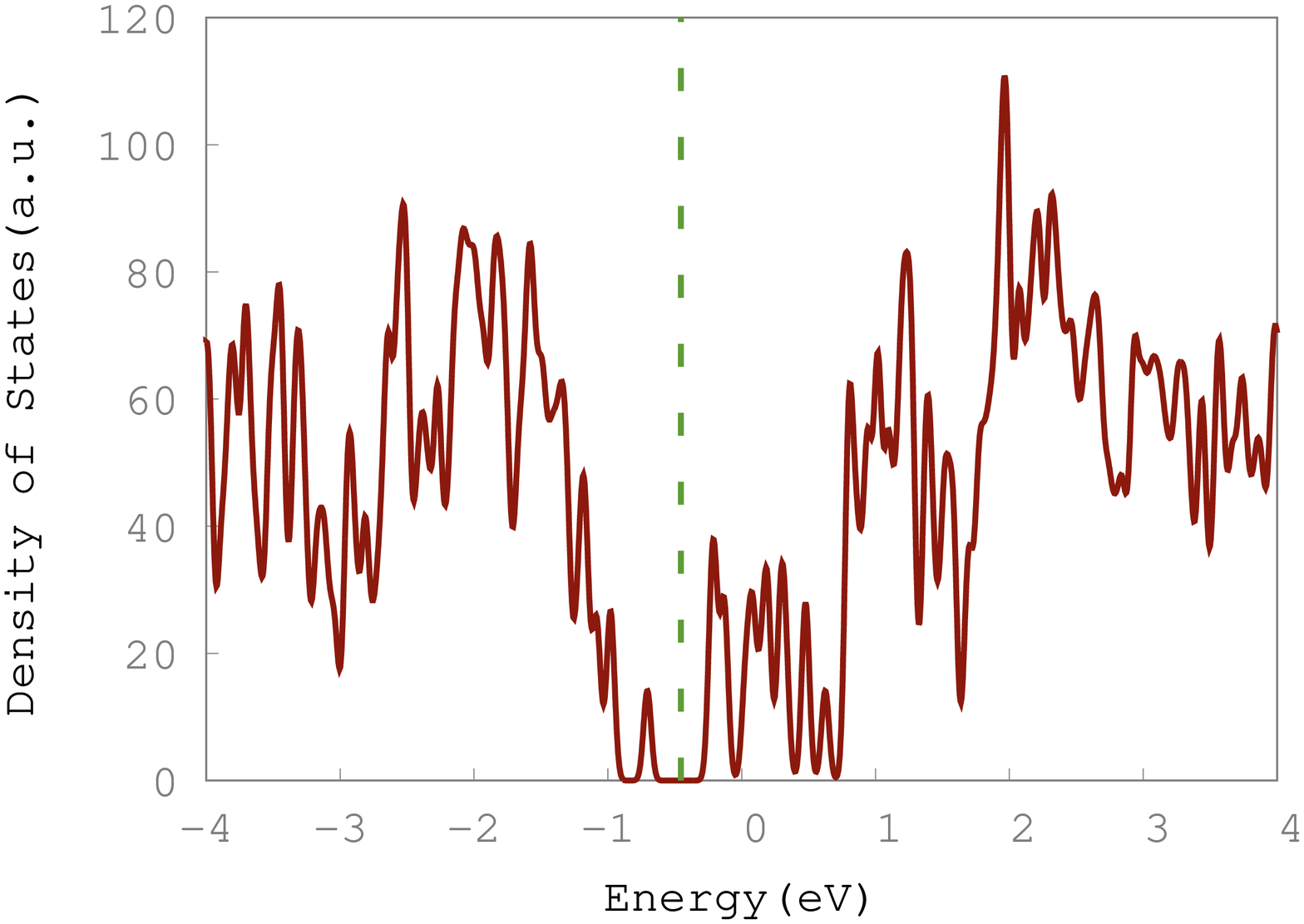}
\caption{PNR 180}
\label{DOS:PNR}
\end{subfigure}
\begin{subfigure}{.45\textwidth}
\includegraphics[width=0.7\textwidth, angle=270]{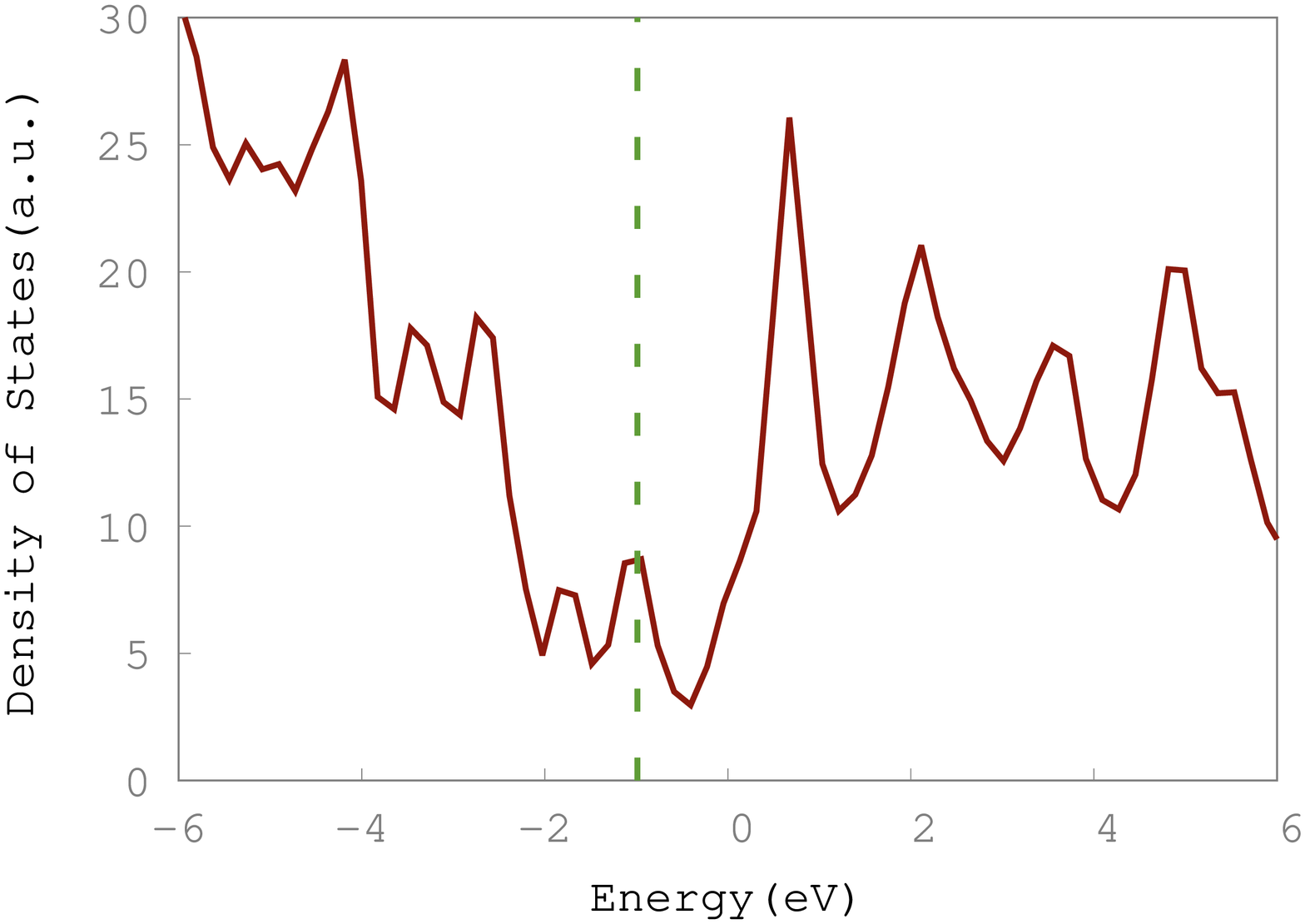}
\caption{GRN 180}
\label{DOS:GRN}
\end{subfigure}
\caption{The total densities of states(DOS) of PNR and GRN 180 atoms, respectively. The fermi levels are marked by the green dash line.
}
\label{DOS}
\end{figure}

Our test systems include a semiconducting phospherene nanoribbon (PNR)
with $180$ atoms, and two metallic graphene (GRN) systems with $180$ and
$6480$ atoms respectively. As shown in Figure ~\ref{DOS}, the GRN system
is a metallic system and PNR is a semi-conductor system with a band gap 
of 0.48 eV. We set the kinetic energy cutoff to be 40
Hartree, and use $15$ adaptive local basis functions per atom.  The size
of the Hamiltonian matrix represented in the adaptive local basis set
for for PNR $180$ and GRN $180$ is $2700$, and is $92700$ for GRN
$6480$, respectively.  All calculations
are carried out on the Edison systems at the National
Energy Research Scientific Computing Center (NERSC). Each node consists
of two Intel ``Ivy Bridge'' processors with $24$ cores in total and
64 gigabyte (GB) of memory. Our implementation only uses MPI.
The number of cores is equal to the number of MPI ranks used in the simulation.

\begin{figure}[H]
\centering
\begin{subfigure}{.45\textwidth}
\includegraphics[width=0.7\textwidth, angle=270]{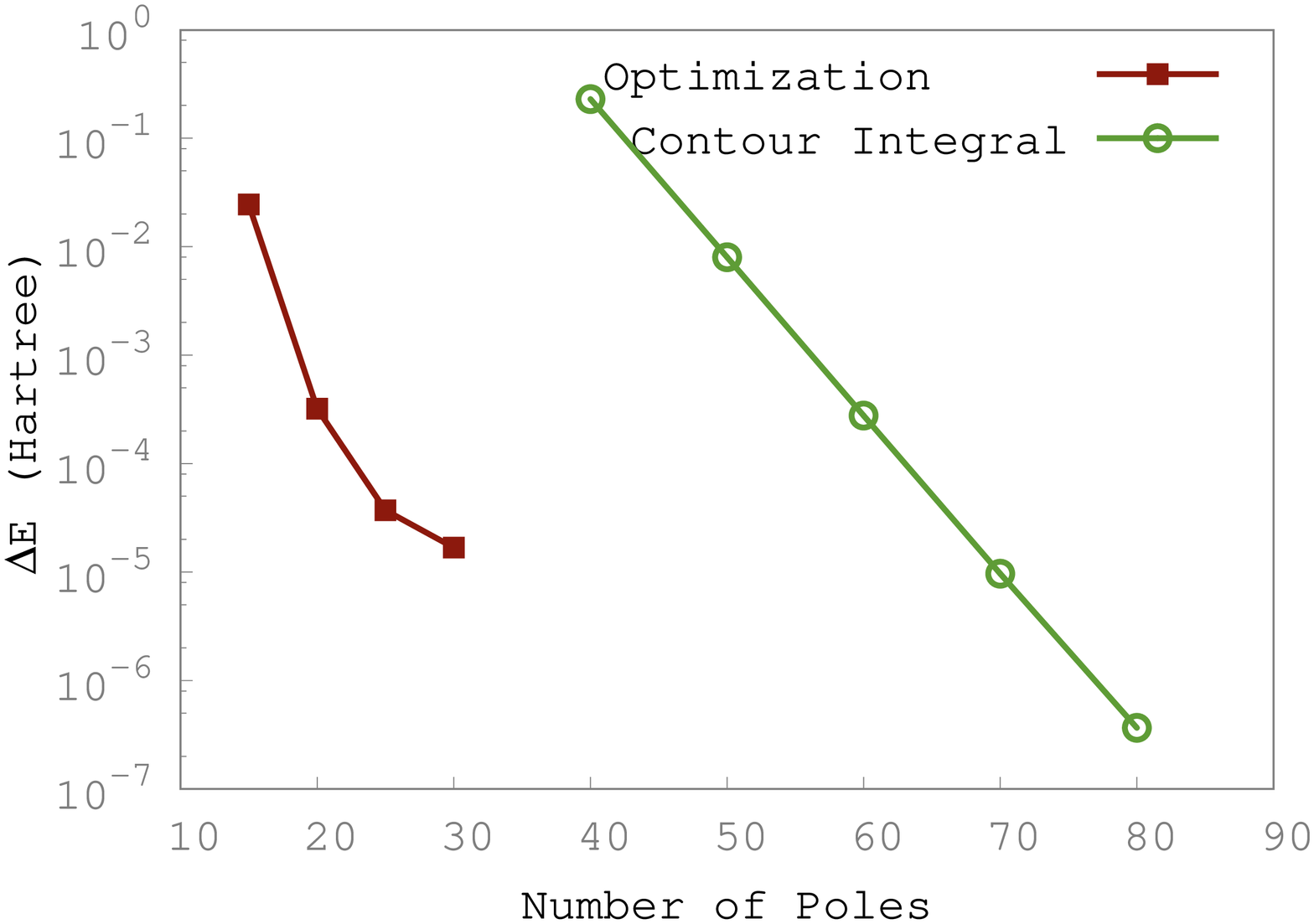}
\caption{Energy}
\label{fig5:sub1}
\end{subfigure}
\begin{subfigure}{.45\textwidth}
\includegraphics[width=0.7\textwidth, angle=270]{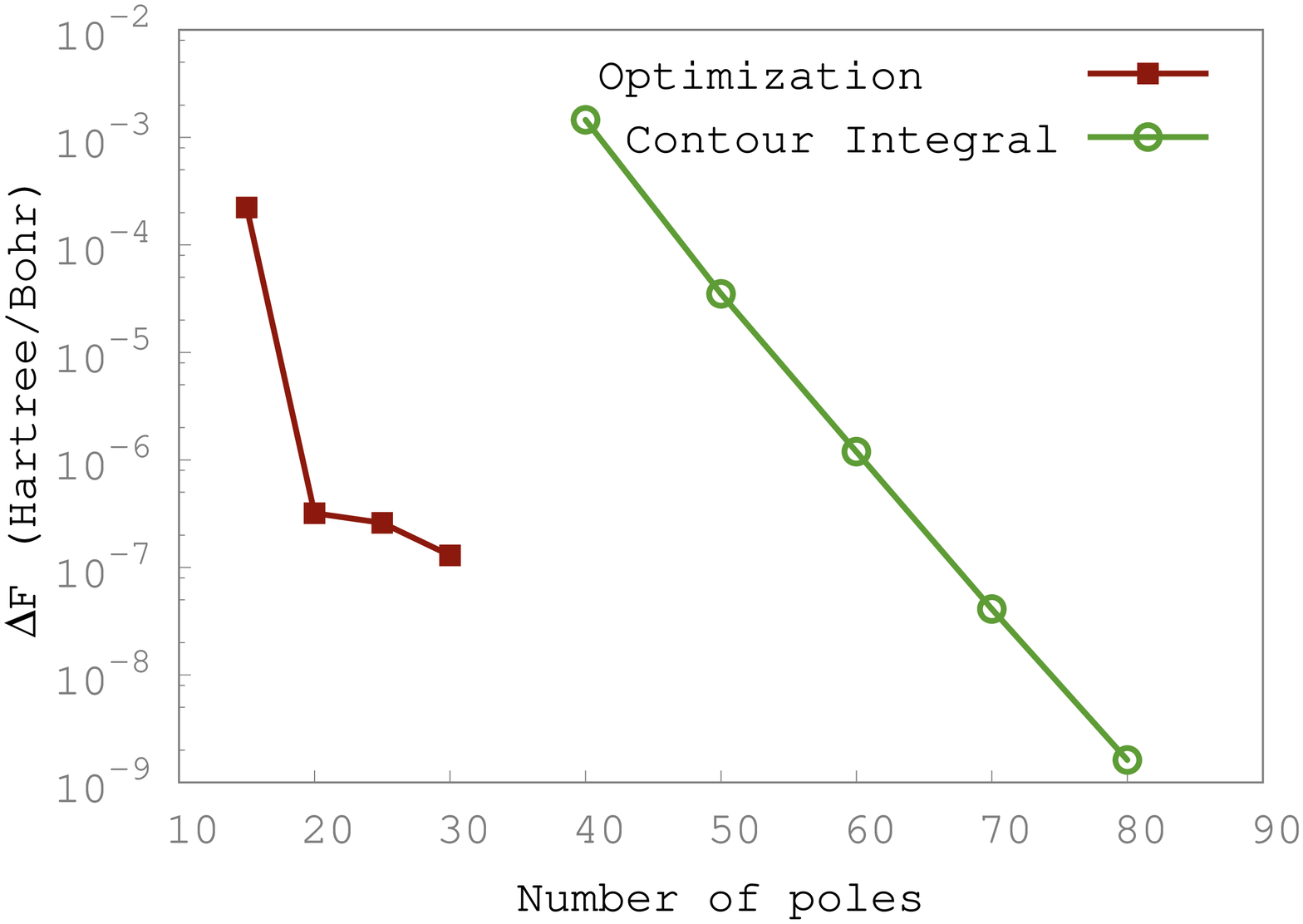}
\caption{Force}
\label{fig5:sub2}
\end{subfigure}
\caption{Error of the energy and force of the PEXSI method with
respect to the number of poles for the graphene 180 system.
}
\label{fig5}
\end{figure}

We first demonstrate the accuracy of the pole expansion using two
approaches: the contour integral approach~\cite{LinLuYingE2009}, and 
the optimization approach by Moussa~\cite{Moussa2016}. The test is performed
using a fully converged Hamiltonian matrix from GRN 180 system,
benchmarked against results from diagonalization methods using the
pdsyevd routine in ScaLAPACK. The accuracy of the pole 
expansion is measured by the error of the energy denoted by $\Delta E$, and 
the maximum error of the force denoted by $\Delta F$,
respectively. As shown in figure \ref{fig5}, in
the contour integral approach, both errors decay exponentially with
respect to the number of poles. When
$80$ poles are used, the pole expansion is highly accurate, as $\Delta E
\approx 10^{-7}$ Hartree and $\Delta F \approx 10^{-9}$ Hartree/Bohr, respectively.  
However, in practical KSDFT calculations, the accuracy reached by
expansion with 60 poles is already sufficiently accurate. 
Such accuracy can be achieved by using 20 poles with the optimization
method, which reduces the number of poles by a factor of $3$.
Nonetheless, when a larger number of poles are requested, the
optimization method solves an increasing more ill-conditioned problem,
and the error stops decreasing after 30 poles are used.

\begin{figure}[H]
\begin{subfigure}{.5\textwidth}
\includegraphics[width=0.7\textwidth,angle=270]{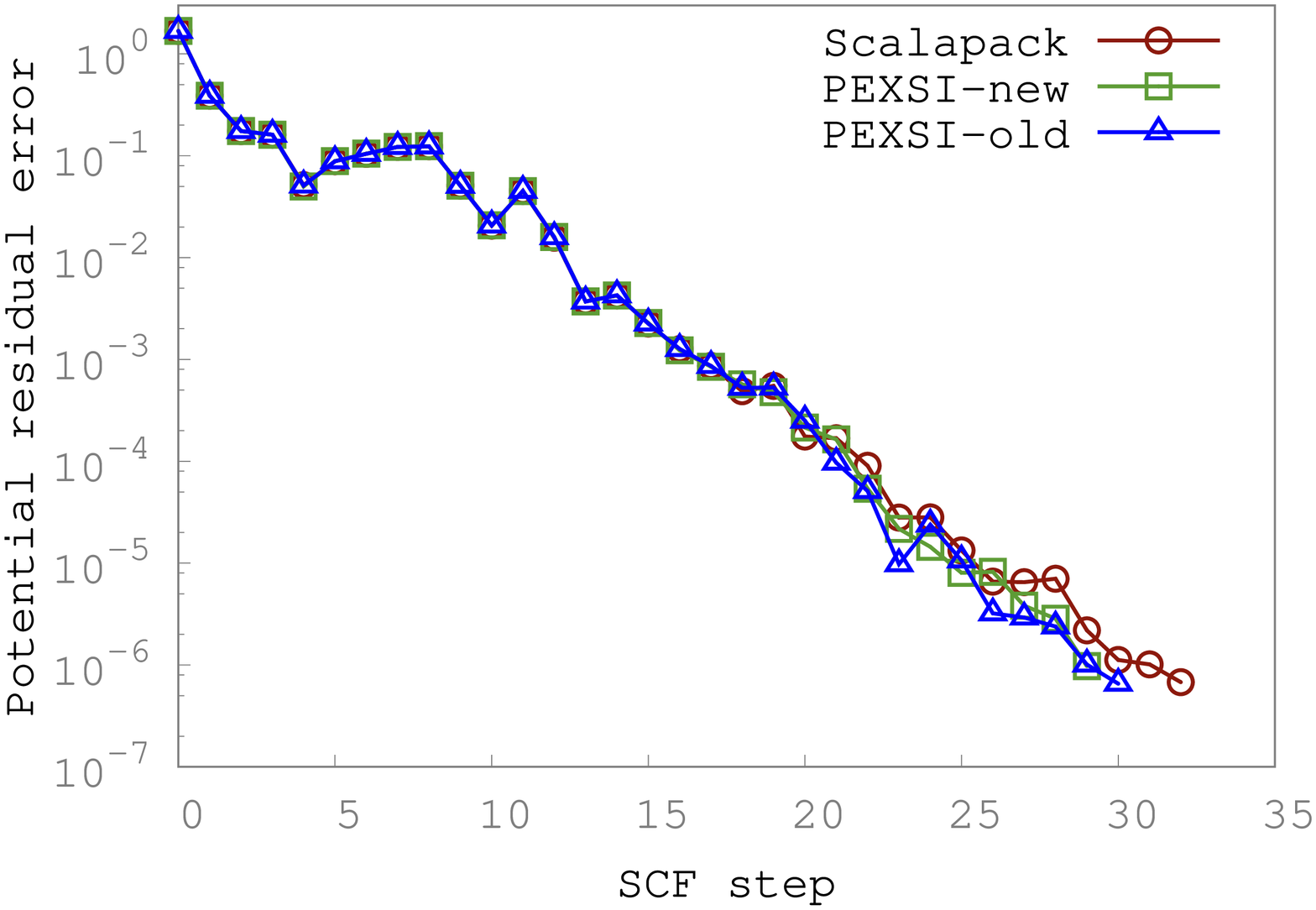}
\caption{PNR 180.}
\label{fig3:sub2}
\end{subfigure}
\begin{subfigure}{.5\textwidth}
\includegraphics[width=0.7\textwidth,angle=270]{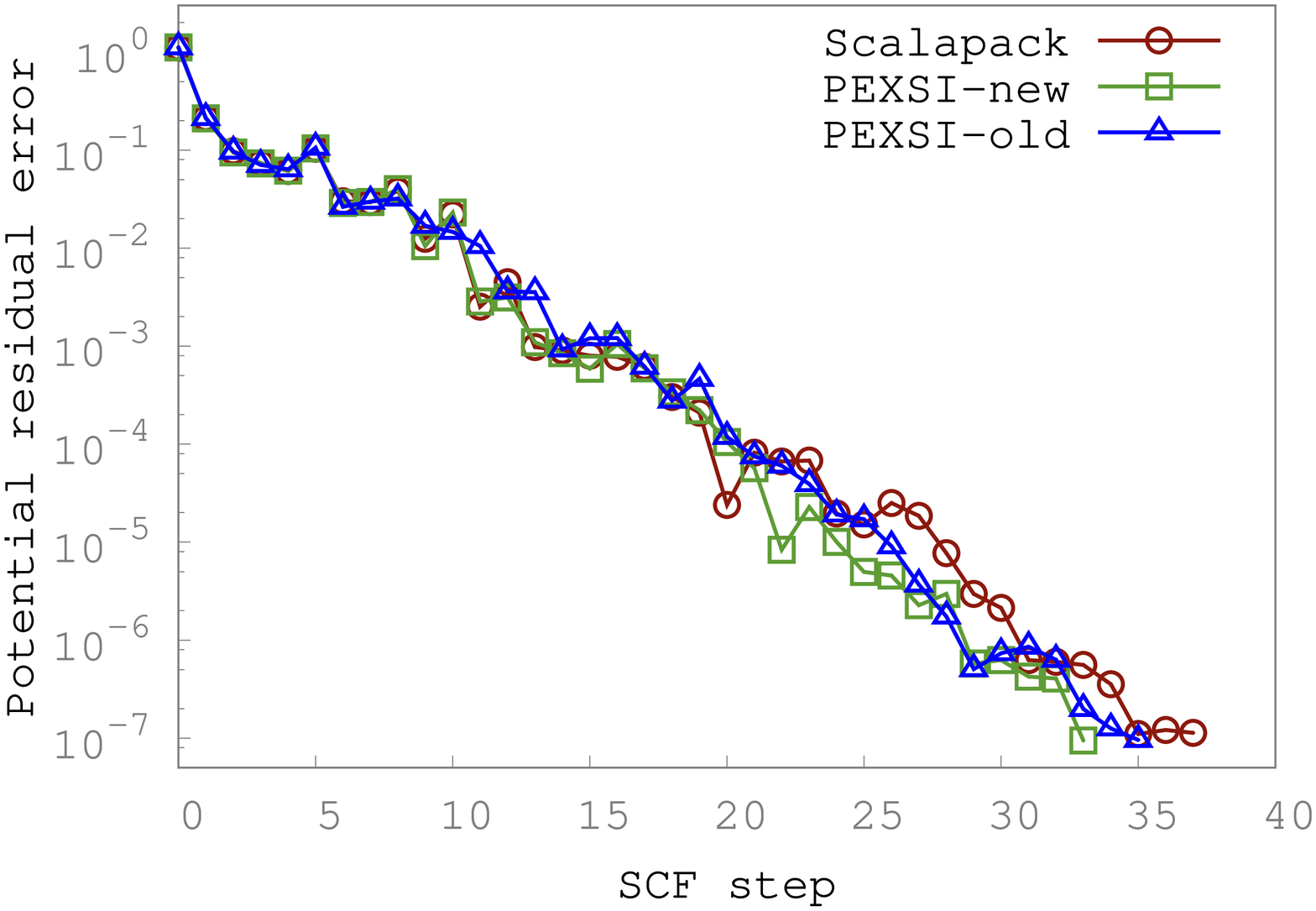}
\caption{GRN 180.}
\label{fig3:sub1}
\end{subfigure}
\caption{SCF convergence with ScaLAPACK and PEXSI-old and PEXSI-new 
along the SCF steps for PNR 180 and GRN 180 systems.
}
\label{fig3}
\end{figure}

As discussed in section~\ref{sec:parallelmu}, the method in
Ref.~\onlinecite{LinGarciaHuhsEtAl2014} uses a hybrid Newton type method
that requires multiple evaluations of the Fermi operator performed sequentially to guarantee convergence
of the chemical potential at each step of the SCF iteration. This is
referred to as the ``PEXSI-old'' method, as opposed to the ``PEXSI-new''
method in this paper.
We then examine the convergence of the SCF iterations using the
PEXSI-new, PEXSI-old, as well as ScaLAPACK methods, respectively. 
The residual error of the potential is defined as
$\norm{V^{\text{out}}-V^{\text{in}}}/\norm{V^{\text{in}}}$, where
$V^{\text{in}}$ and $V^{\text{out}}$ correspond to the input and output
Kohn-Sham potential in each SCF iteration, respectively.
According to Figure~\ref{fig3}, the decay rate of the residual error is
nearly the same for all three methods. In our tests, the convergence
criterion for the residual error is set to $10^{-7}$.  When the SCF
iteration reaches convergence, we find that the error of the total
energy per atom between PEXSI-new and ScaLAPACK is 
$2.8\times10^{-6}$ Hartree/atom for PNR, and is
$5.5\times10^{-8}$ Hartree/atom for GRN, respectively.  Similarly, the
maximum error of the force between PEXSI-new and ScaLAPACK is
$1.48\times10^{-5} \text{Hartree/Bohr}$ for PNR, and is
$2.31\times10^{-6} \text{Hartree/Bohr}$ for GRN, respectively. 


\begin{figure}[H]
\begin{subfigure}{.5\textwidth}
\includegraphics[width=0.7\textwidth, angle=270]{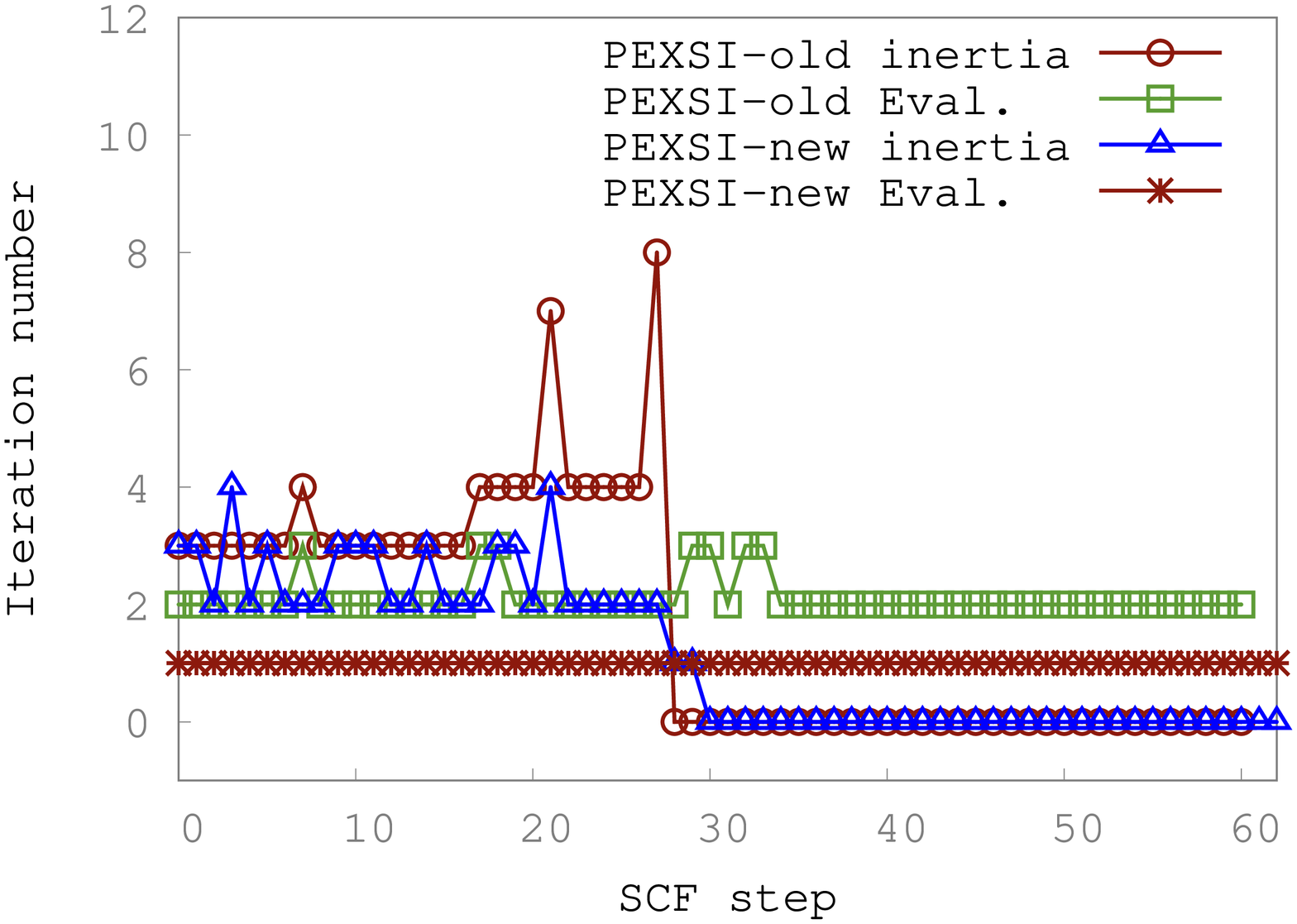}
\caption{Number of inertia counting step and evaluation of the Fermi
operator steps. 
}
\label{fig4:sub1}
\end{subfigure}
\begin{subfigure}{.5\textwidth}
\includegraphics[width=0.7\textwidth, angle=270]{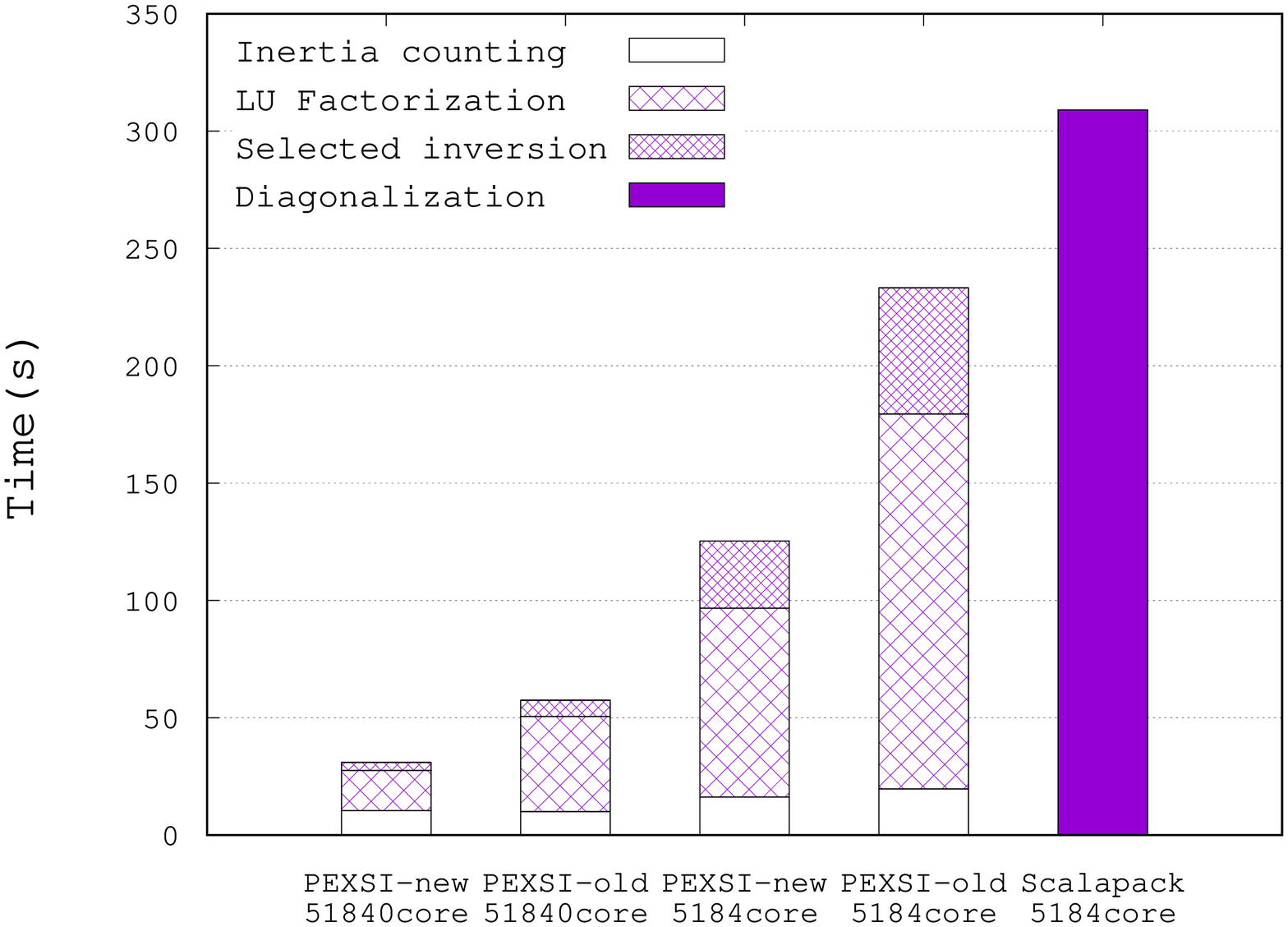}
\caption{The timing for PEXSI-new, PEXSI-old and diagonalization per SCF
step.}
\label{fig4:sub2}
\end{subfigure}
\caption{Comparison of the performance of PEXSI-new, PEXSI-old and
diagonalization for the GRN 6480 system.}
\label{fig4}
\end{figure}

In order to demonstrate the performance of PEXSI for a relatively large
metallic system, we further consider the GRN 6480 atom system.
Figure~\ref{fig4:sub1} shows the number of iterations at the coarse
level (inertia counting) and at the fine level (evaluation of the Fermi
operator) in the PEXSI-new and PEXSI-old methods, respectively.  The
PEXSI-new method uses only one iteration at the fine level by
construction, while the PEXSI-old method uses $2-3$ fine level steps
during the SCF iteration.  We also find that PEXSI-new involves on
average less number of coarse level inertia counting steps as well.
Figure~\ref{fig4:sub2} reports the average wall clock time per SCF
iteration for PEXSI-new, PEXSI-old and diagonalization using ScaLAPACK,
all using 5184 cores.  We also report the timing for LU factorization
and selected inversion separately for the evaluation of the Fermi
operator at the fine level. We can clearly observe that despite the
inertia counting steps at the coarse level may require multiple
iterations, it is much less costly compared to the fine level evaluation
step.  Compared to PEXSI-old,
PEXSI-new reduces the wall clock time by around a factor of $2$ due to
the reduced number of fine level evaluations. We also
remark that while the wall clock time using ScaLAPACK cannot be improved
with further increase of the number of cores, PEXSI can scale up
to 51840 cores, where the average wall clock time per SCF from PEXSI-new
and PEXSI-old becomes 31 seconds and 57 seconds, respectively.




\begin{figure}[H]
\centering
\begin{subfigure}{.45\textwidth}
\includegraphics[width=0.7\textwidth, angle=270]{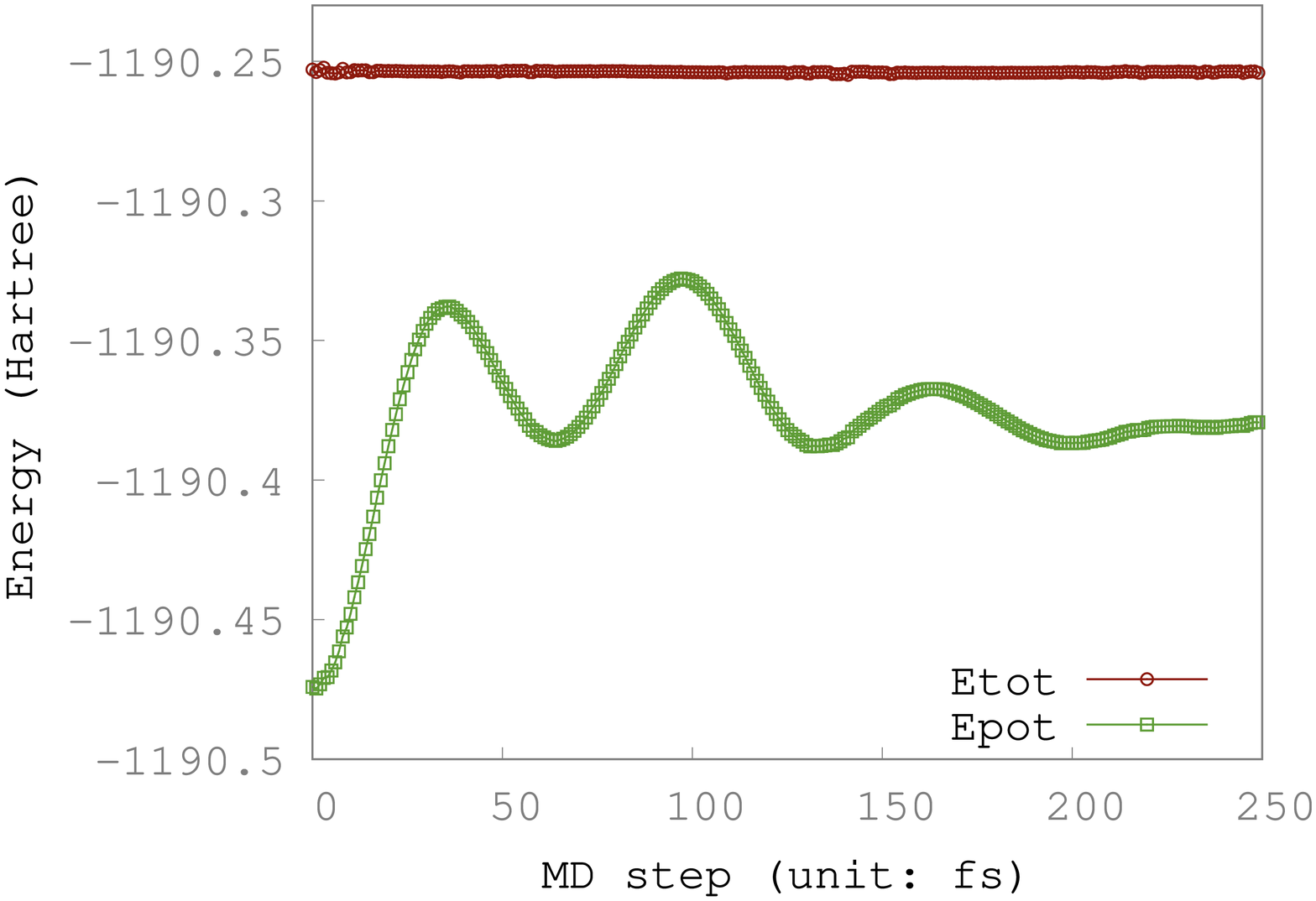}
\caption{Total and potential energy}
\label{fig6:sub1}
\end{subfigure}
\begin{subfigure}{.45\textwidth}
\includegraphics[width=0.7\textwidth,angle=270]{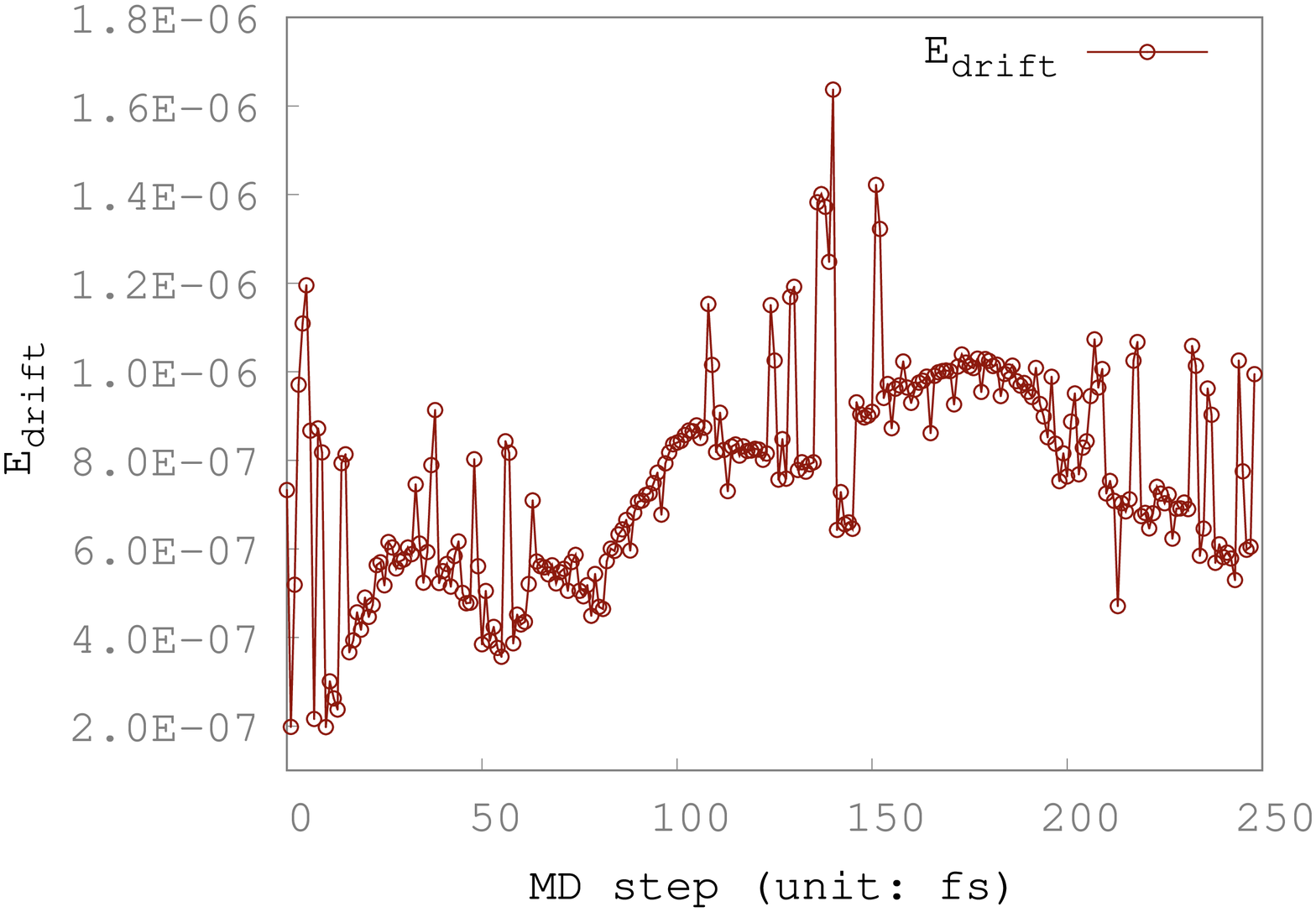}
\caption{Energy drift} 
\label{fig6:sub2}
\end{subfigure}
\caption{The potential energy, total energy and the drift of the total
energy using the PEXSI-new method.}
\label{fig:fig6}
\end{figure}

In order to further demonstrate the effectiveness of the strategy of
on-the-fly convergence of the chemical potential, we apply the PEXSI-new
method to perform \textit{ab initio} molecular dynamics simulation for
PNR 180 system in the NVE ensemble. We use the Verlet method to produce a trajectory of
length $250$ fs.  We set the initial temperature to be 300K. 
Figure~\ref{fig:fig6} shows the potential energy $E_{pot}$ and the total
energy $E_{tot}$, as well as the drift of the total
energy along the simulation, which is defined as
$E_{drift}(t) = \frac{E_{tot}(t) - E_{tot}(0)}{E_{tot}(0)}$.
We find that the use of the PEXSI-new method leads to a very small drift
less than $1.7\times 10^{-6}$. 

\section{Conclusion}\label{sec:conclusion}


In this paper, we developed a robust and efficient approach for finding
the chemical potential in the pole expansion and selected inversion
(PEXSI) approach for solving KSDFT. The main idea of
the new method is not to find the exact chemical potential at each SCF
iteration, but to dynamically update the upper and lower bounds for the
true chemical potential, so that the accuracy of the chemical potential
is comparable to that of the residual error in the SCF iteration. In
particular, when the SCF converges, the chemical potential converges as
well. The new method reduces the number of tunable parameters and is
more robust. In the regime of full
parallelization, the wall clock time in each SCF iteration is always
approximately the same as only one evaluation of the Fermi operator. This reduces the
absolute value as well as the uncertainty of the wall clock time when
the PEXSI method is used for solving KSDFT. 
We demonstrated the efficiency of the new method
using insulating and metallic systems as examples, and the new method is
available in the PEXSI software package. Finally, our approach is not
restricted to the PEXSI method, and can be
applied to other Fermi operator expansion (FOE) type of methods as well.

\section*{Acknowledgement}

This work was partially supported by the National Science Foundation
under Grant No. 1450372 (W. J. and L. L.), by the National Science
Foundation under grant DMS-1652330, the Scientific Discovery through
Advanced Computing (SciDAC) program, and the Center for Applied
Mathematics for Energy Research Applications (CAMERA) funded by U.S.
Department of Energy (L. L.).  This research used resources of the
National Energy Research Scientific Computing Center, a DOE Office of
Science User Facility supported by the Office of Science of the U.S.
Department of Energy under Contract No. DE-AC02-05CH11231. We thank
Jianfeng Lu, Meiyue Shao, Chao Yang and Victor Yu for useful
discussions.

\appendix

\section{Eigenvalue estimates}\label{app:eigenvalue}

Let $\Delta V_{\text{SCF}}(\vrr)$ be the difference of the Kohn-Sham potential
in the real space between two consecutive SCF iterations, and we assume
the basis set denoted by $\Phi$ does not change during the SCF
iterations. The corresponding matrix pencils are $(H,S)$ and
$(\wt{H},S)$, respectively.  The difference of the Hamiltonian matrices
is
\begin{equation}
  \Delta H = \wt{H} - H = \Phi^{T}  \Delta V_{\text{SCF}} \Phi, 
  \label{}
\end{equation}
and the overlap matrix $S=\Phi^{T} \Phi$ remains the same. Denote by $\lambda_{k}$ (resp.
$\wt{\lambda}_{k}$) the $k$-th smallest eigenvalue of $H$ (resp.
$\wt{H}$). Then the Courant-Fisher minimax theorem~\cite{GolubVan2013}
indicates that
\begin{equation}
  \wt{\lambda}_{k} = \min_{\mathrm{dim}\mc{S}=k} \max_{0\ne u\in
  \mc{S}} \left( \frac{u^{T} \wt{H} u}{u^{T} S u} \right),
  \label{}
\end{equation}
where the minimization is over all subspaces $\mc{S}$ with dimension
$k$.
Hence
\begin{equation}
  \begin{split}
  \wt{\lambda}_{k} \le &  \min_{\mathrm{dim}\mc{S}=k} \left[\max_{0\ne u\in
  \mc{S}} \left( \frac{u^{T} H u}{u^{T} S u} \right) + \max_{0\ne u\in
  \mc{S}} \left( \frac{u^{T} \Delta H u}{u^{T} S u} \right)\right]\\\
  \le & \min_{\mathrm{dim}\mc{S}=k} \max_{0\ne u\in
  \mc{S}} \left( \frac{u^{T} H u}{u^{T} S u} \right) + \max_{0\ne u}
  \left( \frac{u^{T} \Delta H u}{u^{T} S u} \right) \\
  \le & \lambda_{k} + \Delta V_{\max}.
  \end{split}
  \label{eqn:lamtupper}
\end{equation}
Here we have used the Courant-Fisher minimax theorem for
$\lambda_{k}$, as well as the estimate of the largest eigenvalue of the matrix
pencil $(\Delta H, S)$ in terms of $\Delta V_{\text{SCF}}$.
Similarly calculation shows
\begin{equation}
  \begin{split}
  \wt{\lambda}_{k} \ge & \min_{\mathrm{dim}\mc{S}=k} \max_{0\ne u\in
  \mc{S}} \left( \frac{u^{T} H u}{u^{T} S u} \right) + \min_{0\ne u}
  \left( \frac{u^{T} \Delta H u}{u^{T} S u} \right) \\
  \ge & \lambda_{k} + \Delta V_{\min}.
  \end{split}
  \label{eqn:lamtlower}
\end{equation}
Again use the fact that the Fermi-Dirac function
$f_{\beta}(\varepsilon)$ is a non-increasing
function, we have
\begin{equation}
  \begin{split}
    \wt{N}_{\beta}(\mu_{\min}+\Delta V_{\min}) := &
    \Tr[f(\wt{\Lambda}-(\mu_{\min}+\Delta V_{\min}) I)]\\
    \le & \Tr[f(\Lambda - \mu_{\min} I)] = N_{\beta}(\mu_{\min}) \le
    N_{e},
  \end{split}
  \label{eqn:Ntlower}
\end{equation}
where the last inequality comes from that $\mu_{\min}$ is a lower bound
for the chemical potential associated with the matrix pencil $(H,S)$.
Similarly
\begin{equation}
  \wt{N}_{\beta}(\mu_{\max}+\Delta V_{\max}) \ge
  N_{\beta}(\mu_{\max}) \ge N_{e}.
  \label{eqn:Ntupper}
\end{equation}
The inequalities~\eqref{eqn:Ntlower} and~\eqref{eqn:Ntupper} establish
that the chemical potential associated with the matrix pencil
$(\wt{H},S)$ is contained in the updated interval $(\mu_{\min}+\Delta
V_{\min},\mu_{\max}+\Delta V_{\max})$.

\bibliographystyle{elsarticle-num}
\bibliography{untitled}

\begin{thebibliography}{10}

\bibitem{ELSI}
\url{http://www.elsi-interchange.org/}.

\bibitem{PEXSI}
\url{http://www.pexsi.org}.

\bibitem{BlumGehrkeHankeEtAl2009}
Volker Blum, Ralf Gehrke, Felix Hanke, Paula Havu, Ville Havu, Xinguo Ren,
  Karsten Reuter, and Matthias Scheffler.
\newblock Ab initio molecular simulations with numeric atom-centered orbitals.
\newblock {\em Computer Physics Communications}, 180(11):2175--2196, 2009.

\bibitem{BowlerMiyazakiGillan2002}
DR~Bowler, T~Miyazaki, and MJ~Gillan.
\newblock Recent progress in linear scaling ab initio electronic structure
  techniques.
\newblock {\em Journal of Physics: Condensed Matter}, 14(11):2781, 2002.

\bibitem{BowlerMiyazaki2012}
DR~Bowler and Tsuyoshi Miyazaki.
\newblock Methods in electronic structure calculations.
\newblock {\em Reports on Progress in Physics}, 75(3):036503, 2012.

\bibitem{ATK}
Mads Brandbyge, Jos{\'e}-Luis Mozos, Pablo Ordej{\'o}n, Jeremy Taylor, and Kurt
  Stokbro.
\newblock Density-functional method for nonequilibrium electron transport.
\newblock {\em Physical Review B}, 65(16):165401, 2002.

\bibitem{FattebertBernholc2000}
J-L Fattebert and J~Bernholc.
\newblock Towards grid-based o (n) density-functional theory methods: Optimized
  nonorthogonal orbitals and multigrid acceleration.
\newblock {\em Physical Review B}, 62(3):1713, 2000.

\bibitem{Goedecker1993}
S~Goedecker.
\newblock Integral representation of the fermi distribution and its
  applications in electronic-structure calculations.
\newblock {\em Physical Review B}, 48(23):17573, 1993.

\bibitem{Goedecker1999}
Stefan Goedecker.
\newblock Linear scaling electronic structure methods.
\newblock {\em Reviews of Modern Physics}, 71(4):1085, 1999.

\bibitem{GolubVan2013}
GH~Golub and C~Van~Loan.
\newblock Matrix computations, 4th.
\newblock {\em Johns Hopkins}, 2013.

\bibitem{HineHaynesMostofiEtAl2009}
Nicholas~DM Hine, Peter~D Haynes, Arash~A Mostofi, C-K Skylaris, and Mike~C
  Payne.
\newblock Linear-scaling density-functional theory with tens of thousands of
  atoms: Expanding the scope and scale of calculations with onetep.
\newblock {\em Computer Physics Communications}, 180(7):1041--1053, 2009.

\bibitem{HuLinYang2015a}
Wei Hu, Lin Lin, and Chao Yang.
\newblock Dgdft: A massively parallel method for large scale density functional
  theory calculations.
\newblock {\em The Journal of chemical physics}, 143(12):124110, 2015.

\bibitem{HuLinYangEtAl2016}
Wei Hu, Lin Lin, Chao Yang, Jun Dai, and Jinlong Yang.
\newblock Edge-modified phosphorene nanoflake heterojunctions as highly
  efficient solar cells.
\newblock {\em Nano letters}, 16(3):1675--1682, 2016.

\bibitem{HuLinYangEtAl2014}
Wei Hu, Lin Lin, Chao Yang, and Jinlong Yang.
\newblock Electronic structure and aromaticity of large-scale hexagonal
  graphene nanoflakes.
\newblock {\em The Journal of chemical physics}, 141(21):214704, 2014.

\bibitem{JacquelinLinWichmannEtAl2016}
Mathias Jacquelin, Lin Lin, Nathan Wichmann, and Chao Yang.
\newblock Enhancing the scalability and load balancing of the parallel selected
  inversion algorithm via tree-based asynchronous communication.
\newblock {\em arXiv preprint arXiv:1504.04714}, 2016.

\bibitem{JacquelinLinYang2016}
Mathias Jacquelin, Lin Lin, and Chao Yang.
\newblock Pselinv—a distributed memory parallel algorithm for selected
  inversion: The symmetric case.
\newblock {\em ACM Transactions on Mathematical Software (TOMS)}, 43(3):21,
  2016.

\bibitem{Kohn1996}
Walter Kohn, Axel~D Becke, and Robert~G Parr.
\newblock Density functional theory of electronic structure.
\newblock {\em The Journal of Physical Chemistry}, 100(31):12974--12980, 1996.

\bibitem{LiNunesVanderbilt1993}
X-P Li, RW~Nunes, and David Vanderbilt.
\newblock Density-matrix electronic-structure method with linear system-size
  scaling.
\newblock {\em Physical Review B}, 47(16):10891, 1993.

\bibitem{LiDemmel2003}
X.S. Li and J.W. Demmel.
\newblock {SuperLU\_DIST}: A scalable distributed-memory sparse direct solver
  for unsymmetric linear systems.
\newblock {\em ACM Trans. Math. Software}, 29:110, 2003.

\bibitem{LinChenYangEtAl2013}
Lin Lin, Mohan Chen, Chao Yang, and Lixin He.
\newblock Accelerating atomic orbital-based electronic structure calculation
  via pole expansion and selected inversion.
\newblock {\em Journal of Physics: Condensed Matter}, 25(29):295501, 2013.

\bibitem{LinGarciaHuhsEtAl2014}
Lin Lin, Alberto Garc{\'\i}a, Georg Huhs, and Chao Yang.
\newblock Siesta-pexsi: massively parallel method for efficient and accurate ab
  initio materials simulation without matrix diagonalization.
\newblock {\em Journal of Physics: Condensed Matter}, 26(30):305503, 2014.

\bibitem{LinLuYingE2009}
Lin Lin, Jianfeng Lu, Lexing Ying, Roberto Car, E~Weinan, et~al.
\newblock Fast algorithm for extracting the diagonal of the inverse matrix with
  application to the electronic structure analysis of metallic systems.
\newblock {\em Communications in Mathematical Sciences}, 7(3):755--777, 2009.

\bibitem{LinLuYingE2012}
Lin Lin, Jianfeng Lu, Lexing Ying, and E~Weinan.
\newblock Optimized local basis set for kohn--sham density functional theory.
\newblock {\em Journal of Computational Physics}, 231(13):4515--4529, 2012.

\bibitem{LinYangMezaEtAl2011}
Lin Lin, Chao Yang, Juan~C Meza, Jianfeng Lu, Lexing Ying, et~al.
\newblock Selinv---an algorithm for selected inversion of a sparse symmetric
  matrix.
\newblock {\em ACM Transactions on Mathematical Software (TOMS)}, 37(4):40,
  2011.

\bibitem{McWeeny1960}
Rev McWeeny.
\newblock Some recent advances in density matrix theory.
\newblock {\em Reviews of Modern Physics}, 32(2):335, 1960.

\bibitem{MohrRatcliffBoulangerEtAl2014}
Stephan Mohr, Laura~E Ratcliff, Paul Boulanger, Luigi Genovese, Damien Caliste,
  Thierry Deutsch, and Stefan Goedecker.
\newblock Daubechies wavelets for linear scaling density functional theory.
\newblock {\em The Journal of chemical physics}, 140(20):204110, 2014.

\bibitem{Moussa2016}
Jonathan~E Moussa.
\newblock Minimax rational approximation of the fermi-dirac distribution.
\newblock {\em The Journal of chemical physics}, 145(16):164108, 2016.

\bibitem{ProdanKohn2005}
Emil Prodan and Walter Kohn.
\newblock Nearsightedness of electronic matter.
\newblock {\em Proceedings of the National Academy of Sciences of the United
  States of America}, 102(33):11635--11638, 2005.

\bibitem{SolerArtachoGaleEtAl2002}
Jos{\'e}~M Soler, Emilio Artacho, Julian~D Gale, Alberto Garc{\'\i}a, Javier
  Junquera, Pablo Ordej{\'o}n, and Daniel S{\'a}nchez-Portal.
\newblock The siesta method for ab initio order-n materials simulation.
\newblock {\em Journal of Physics: Condensed Matter}, 14(11):2745, 2002.

\bibitem{Sylvester1852}
James~Joseph Sylvester.
\newblock Xix. a demonstration of the theorem that every homogeneous quadratic
  polynomial is reducible by real orthogonal substitutions to the form of a sum
  of positive and negative squares.
\newblock {\em Philosophical Magazine Series 4}, 4(23):138--142, 1852.

\bibitem{VandeVondeleKrackMohamedEtAl2005}
Joost VandeVondele and Michiel Sprik.
\newblock A molecular dynamics study of the hydroxyl radical in solution
  applying self-interaction-corrected density functional methods.
\newblock {\em Physical Chemistry Chemical Physics}, 7(7):1363--1367, 2005.

\bibitem{Wolberg1999}
George Wolberg and Itzik Alfy.
\newblock Monotonic cubic spline interpolation.
\newblock In {\em cgi}, page 188. IEEE, 1999.

\bibitem{Yang1991}
Weitao Yang.
\newblock Direct calculation of electron density in density-functional theory.
\newblock {\em Physical Review Letters}, 66(11):1438, 1991.

\end{thebibliography}

\end{document}